\documentclass[a4paper,12pt,numbers,sort&compress]{article}
\pdfoutput=1
\usepackage{amsfonts,amsmath,amssymb}
\usepackage[paper=letterpaper,margin=1.0in]{geometry}
\usepackage{graphicx}
\usepackage{cite}
\usepackage{hyperref}

\newcommand{\be}{\begin{equation}}
\newcommand{\ee}{\end{equation}}
\newcommand{\bea}{\begin{eqnarray}}
\newcommand{\eea}{\end{eqnarray}}
\newcommand{\ba}{\begin{array}}
\newcommand{\ea}{\end{array}}
\newcommand{\ben}{\begin{enumerate}}
\newcommand{\een}{\end{enumerate}}
\newcommand{\bi}{\begin{itemize}}
\newcommand{\ei}{\end{itemize}}
\newcommand{\bc}{\begin{center}}
\newcommand{\ec}{\end{center}}
\newcommand{\bfig}{\begin{figure}}
\newcommand{\efig}{\end{figure}}
\newcommand{\bq}{\begin{quotation}}
\newcommand{\eq}{\end{quotation}}
\newcommand{\bt}{\begin{table}}
\newcommand{\et}{\end{table}}
\newcommand{\btab}{\begin{tabular}}
\newcommand{\etab}{\end{tabular}}
\newcommand{\bs}{\begin{slide}}
\newcommand{\es}{\end{slide}}
\newcommand{\nn}{\nonumber}
\newcommand{\eref}[1]{(\ref{#1})}
\newcommand{\todo}[1]{{\bf ?????!!!! #1 ?????!!!!}\marginpar{$\Longleftarrow$}}

\newcommand{\setall}{\setcounter{equation}{0}}

\newcommand{\comment}[1]{}
\newcommand{\mC}{\mathbb{C}}

\newcommand{\bmQ}{\overline{\mathbb{Q}}}
\newcommand{\mZ}{\mathbb{Z}}
\newcommand{\mT}{\mathbb{T}}
\newcommand{\mP}{\mathbb{P}}

\begin{document}

${}$

\vspace{-0.6in}

{} ~ {} ~ {} ~ {} ~ {} ~ \hbox{\small IMPERIAL-TP-11-AH-04, QMUL-PH-11-04}
\break

\vspace{0.2in}

\centerline{{\Huge \bf The Beta Ansatz:}}
\centerline{{${}$}}
\centerline{{\large \bf A Tale of Two Complex Structures}}
\medskip

\vspace{.4cm}

\centerline{
{\large Amihay Hanany}$^1$,
{\large Yang-Hui He}$^2$,
{\large Vishnu Jejjala}$^3$,
{\large Jurgis Pasukonis}$^3$,
}
\vspace{.4cm}
\centerline{
{\large Sanjaye Ramgoolam}$^3$,
{\large and}
{\large Diego Rodriguez-Gomez}$^4$
\footnote{
a.hanany@imperial.ac.uk,
yang-hui.he.1@city.ac.uk,
v.jejjala@qmul.ac.uk,
j.pasukonis@qmul.ac.uk,
s.ramgoolam@qmul.ac.uk,
drodrigu@physics.technion.ac.il}
}

\vspace*{3.0ex}

\begin{center}
{${}^{1}$ Theoretical Physics Group, The Blackett Laboratory,\\
Imperial College, Prince Consort Road, London SW7 2AZ, UK\\
}
\vspace*{1.5ex}
{${}^{2}$ Department of Mathematics, City University, London,\\
Northampton Square, London EC1V 0HB, UK;\\
School of Physics, NanKai University, Tianjin, 300071, P.R.~China;\\
Merton College, University of Oxford, OX14JD, UK\\
}
\vspace*{1.5ex}
{${}^{3}$ Department of Physics, Queen Mary, University of London,\\
Mile End Road, London E1 4NS, UK\\
}
\vspace*{1.5ex}
{${}^{4}$ Department of Physics, Technion, Haifa, 3200, Israel;\\}
{Department of Mathematics and Physics,\\
University of Haifa at Oranim, Tivon, 36006, Israel\\
}
\end{center}

\vspace*{4.0ex}
\centerline{\textbf{Abstract}} \bigskip
Brane tilings, sometimes called dimer models, are a class of bipartite graphs on a torus which encode the gauge theory data of four-dimensional SCFTs dual to D$3$-branes probing toric Calabi--Yau threefolds.
An efficient way of encoding this information exploits the theory of \textit{dessin d'enfants}, expressing the structure in terms of a permutation triple, which is in turn related to a Belyi pair, namely a holomorphic map from a torus to a $\mathbb{P}^1$ with three marked points.
The procedure of $a$-maximization, in the context of isoradial embeddings
of the dimer, also associates a complex structure to the torus, determined by the $R$-charges in the SCFT, which can be compared with the Belyi complex structure.
Algorithms for the explicit construction of the Belyi pairs are described in detail.
In the case of orbifolds, these algorithms are related to the construction of covers of elliptic curves, which exploits the properties of Weierstra\ss\ elliptic functions.
We present a counterexample to a previous conjecture identifying the complex structure of the Belyi curve to the complex structure associated with $R$-charges.

\newpage

\setcounter{footnote}{0}

\tableofcontents

\newpage

\section{Introduction}
\label{introduction}\setall

The low-energy physics of D$3$-branes probing a toric Calabi--Yau threefold conical singularity $X$ is given in terms of a four-dimensional conformal field theory with four supercharges.
Through the AdS/CFT duality~\cite{Maldacena:1997re, Gubser:1998bc, Witten:1998qj}, these superconformal field theories (SCFTs) are dual to Type IIB superstring theory on $AdS_5\times \mathcal{B}$, where $\mathcal{B}$ is the base of $X$ seen as a cone $\mathbb{R}^+\times \mathcal{B}$.
This is by now a well known story.
Quite remarkably, these theories --- including the archetypal ${\cal N}=4$ super-Yang--Mills --- can each be encoded in a bipartite graph drawn on a torus.
This is called a \textit{dimer model} or, in a more stringy language, a \textit{brane tiling}~\cite{Hanany:2005ve,Franco:2005rj}.
A nice interpretation of this graph is given in the mirror Type IIA background, as described in~\cite{Feng:2005gw}, by a so-called alga projection.
Moreover, it is also possible to relate the setup to a certain fivebrane system~\cite{Franco:2005rj, Imamura:2007dc}, which generalizes the brane box~\cite{Hanany:1997tb, Hanany:1998it} and brane diamond~\cite{Aganagic:1999fe} constructions, whereby giving rise to a \textit{brane tiling}.
Introductions to dimer models and brane tilings may be found in the reviews~\cite{Kennaway:2007tq, Yamazaki:2008bt}.

Of late, it was observed in~\cite{Jejjala:2010vb} that dimers,
regarded as a bipartite graphs on a torus $\mathbb{T}^2$, can naturally
 be interpreted in terms of Grothendieck's \textit{dessins d'enfants}~\cite{grotesquisse},
 or children's drawings.\footnote{
Dessins and Belyi pairs have also appeared in string
 theory in the context of Seiberg--Witten curves for $\mathcal{N}=2$
 theories~\cite{Ashok:2006br} and Matrix Models~\cite{BI, Looijenga,
 Koch:2010zz, Gopakumar:2011ev}.}
By labeling each edge of the dimer with a number, one encodes the data of the graph in terms of three permutation elements in the symmetric group $S_d$ on $d$ elements.
Here, $d$, which is the number of edges, corresponds to the number of fields in the dual SCFT by virtue of the standard dimer model rules.
By the Riemann existence theorem (see, for example,~\cite{lanzvon})
 the combinatorial data of the dimer
determines a unique holomorphic map $\beta$ (up to equivalence
under holomorphic reparameterizations of the curve) from the
torus $\mathbb{T}^2$ to $\mathbb{P}^1$, with branch points at $\{0,\,1,\,\infty\}$. Henceforth, as is common in the literature but not universal, we refer to the three special points $\{0,\,1,\,\infty\}$ on the $\mathbb{P}^1$ as branch
 points and their pre-images on ${\mT}^2$, where the derivative of $\beta $ vanishes, as ramification points.

Such maps have attracted much attention in the mathematical literature since, due to an important result by G.~V.~Belyi~\cite{belyi}, their existence implies that the Riemann surface on which they are defined --- in this case a torus --- can be defined over $\overline{\mathbb{Q}}$, the field of algebraic numbers.
Thus the \textbf{Belyi pair}, consisting of (1) the Riemann surface $\mathbb{T}^2$
which is the source of the Belyi map along with
(2) the holomorphic map $\beta$, acquires a special importance.

Explicit constructions of Belyi pairs are difficult, in particular due to the rigidity of the construction, which allows no moduli.
Indeed, whereas ramified maps from $\mathbb{P}^1$ to $\mathbb{P}^1$ have algorithmic methods of construction~\cite{schneps1994dessins}, Belyi maps from $\mathbb{T}^2$ to $\mathbb{P}^1$ has so far defied a general explicit treatment~\cite{zapponi1997dessins, couveignes1994dessins}.
It was observed~\cite{Jejjala:2010vb}, in the context of constructing Belyi pairs associated to orbifolds of Calabi--Yaus, that an
 infinite series of pairs can be constructed from a ``parent map''
 by considering the map on the $n$-fold unbranched cover of the
 original $\mathbb{T}^2$. The field theory
construction corresponding to orbifolds is based on~\cite{Douglas:1996sw}. The relation between orbifolds and
$n$-fold covers of tori has been previously observed and
explored, in the pure dimer related context,
by~\cite{Hanany:2010cx, Davey:2010px, Hanany:2010ne, newpaper}.
In this note, we will give general
constructions for the $n$-fold unbranched covers,
and apply the constructions to give new explicit examples of Belyi pairs
for small $n$.

In the course of exploring the meaning of the Belyi pair
associated to a Calabi--Yau and associated SCFT, a first
step is to explore the most basic geometrical structure
associated with the Belyi pair, namely the complex structure
denoted $\tau_B$ on the Belyi curve, which makes the map $\beta $
holomorphic. It is known that $R$-charges can be associated to
angles of the dimer in the isoradial construction of
 dimers~\cite{Hanany:2005ss}.
The $R$-charges of the SCFT, determined by $a$-maximization
fix the structure of the dimer, hence its periodicity.
 This determines a complex structure on the torus which
supports the dimer, denoted by $\tau_R$, which was
highlighted in~\cite{Jejjala:2010vb}.
For the case of the conifold and $\mC^3$ and their orbifolds,
 this complex structure $\tau_R$ was shown to agree with $\tau_B$.
It was conjectured that this equality holds generally.

From a physical point of view, a natural class of SCFTs to consider
after orbifolds, are the other toric phases which can be reached by means of Seiberg~\cite{Seiberg:1994pq}, or toric, dualities~\cite{Feng:2000mi, Feng:2001xr}.
Below, we study such phases in the particular examples of the conifold and its orbifolds. We find that, in one case, the equality $ \tau_B = \tau_R $
extends beyond its prediction by orbifolding the conifold.
However, we also find a counterexample in the context of toric phases
related to orbifolds, to the conjectured equality $ \tau_B = \tau_R $.
The relations between $\tau_B$ and $\tau_R$ are
thus more intricate and require a deeper physical explanation.
In the course of these investigations, we found a proof
that $\tau_R$ is invariant under Seiberg dualities.
This forms part of a forthcoming work~\cite{Seibergdualtaus}
on invariants.

The structure of this note is as follows.
In Section~\ref{review} we give a lightning review of the combinatorial description of dimers and its relation with holomorphic maps from $\mathbb{T}^2$ into $\mathbb{P}^1$.
In Section~\ref{algorithm} we discuss a general algorithm for constructing Belyi pairs. The discussion separates a class of cases which
reduces, subject to specified conditions on the structure of the dimer,
 to the simpler problem of Belyi maps from $ \mP^1$ to $\mP^1$ (Appendix~\ref{P1belyi}).
In Section~\ref{orbifolds} we describe, following~\cite{Jejjala:2010vb}, how orbifolds are constructed in terms of the Belyi pair.
Thus prepared, in Section~\ref{chicha}, we introduce a general procedure to explicitly construct such covers.
With this newly developed technology at hand, we study orbifolds of $\mathbb{C}^3$ and the conifold in Sections~\ref{C3example} and~\ref{Conifoldexample}.
Then, in Section~\ref{Seibergduality} we explore different phases of the orbifolds obtained, through Seiberg duality in the field theory.
In particular, we find a counterexample to the conjectured equivalence of $\tau_B$ and $\tau_R$.
We finish in Section~\ref{conclusions} with some concluding remarks and mention some open problems for future research.

\section{Belyi pairs and dimer models}
\label{review}\setall

In this section, we quickly summarize the rudiments of the dimer model (equivalently, the brane tiling) representation of gauge theories arising from branes at toric singularities, as well as the recent realization that this can also be interpreted as a dessin d'enfant, and hence be encoded by a Belyi pair.
A dimer model is a bipartite graph, \textit{i.e.}, consisting of two sets of nodes, say black and white, such that only nodes of opposite color are allowed to be connected by an undirected edge.
This finite graph is then drawn on a torus, hence constituting a periodic tiling of the plane.
From empirical observation of field theories, we restrict to the case of \textit{balanced} bipartite graphs, for which we have an equal number of black and white nodes.

The dimer captures the information of the dual field
 theory in the following way.
The faces represent $U(N)$ gauge group factors while its edges represent fields in the bifundamental representation of the two faces which the edge separates.
The orientation of the torus on which the graph is embedded distinguishes the fundamental representation from the antifundamental representation.
The reader may recognize this as the dual graph manifestation of a periodic quiver tiling.
The advantage of the dimer model is that
it also compactly encodes the superpotential.
Monomial terms are formed as ordered strings of edges (fields) going, say, clockwise around white vertices and anti-clockwise around black vertices.
Thereby, the superpotential is reproduced by adding together such monomials with plus sign for those originating from white nodes and a minus sign for those originating from black nodes.
There is a catalog of all the known dimer models thus far, and we refer the reader to~\cite{Davey:2009bp}.

As briefly mentioned in the introduction, the dimer model can be fully encoded by a set of permutations and subsequently, by a Belyi pair, as we now recall from~\cite{Jejjala:2010vb}.
We first label each of the $d$ edges with a number from $1$ to $d$.
Then, we construct a string of numbers, dubbed~\textbf{cycles}, associated to each black node by going, say, anti-clockwise around each node.
Adjoining all such cycles for each vertex gives an element, in the standard cycle notation, of the symmetric group $S_d$ on $d$ numbers, which we will denote as $\sigma_B$.
By going anti-clockwise around the
 white nodes we obtain the permutation element $\sigma_W$.
We traverse the white nodes in the same orientation as we have traversed the black nodes.\footnote{
We can view this prescription
as traversing the edges around all the vertices according to
the orientation of the $\mathbb{T}^2$ where the dimer lives.
This way of constructing the permutations allows one to read off
the genus of the torus from the cycle structure of the three
permutations~\cite{Jejjala:2010vb} using the
Riemann--Hurwitz formula. Note that the superpotential
terms are read off from $ {\sigma}_B , {\sigma}_W^{-1} $.
Exchanging the roles of ${\sigma}_W $ and
$\sigma_W^{-1} $, is related to { \it untwisting} the dimer
~\cite{Stienstra:2007dy,Feng:2005gw} which can produce Riemann surfaces
of genus not equal to one.}
We stress that both types of nodes are circled
with \textit{the same orientation}.

Therefore, we have two strings of numbers
defining the cycles of permutations
 $\sigma_B,\,\sigma_W$ in $S_d$.
We can naturally form a third permutation $\sigma_{\infty}$ by demanding the Calabi--Yau condition, with multiplication in $S_d$, that
\begin{equation}
\label{relation}
\sigma_B\cdot\sigma_W\cdot\sigma_{\infty}=1 ~.
\end{equation}
The cycles in the third permutation $\sigma_{\infty}$ are associated to the faces of the dimer (\textit{i.e.}, the gauge groups in the SCFT).
We refer the reader to~\cite{Jejjala:2010vb} for further details and examples.

Writing the dimer in the above language lends itself well to the interpretation of Belyi maps.
First,~(\ref{relation}) coincides with the relation among the homology generators on a $\mathbb{P}^1$ marked with three points.
In fact, the set of permutations $\{\sigma_B,\,\sigma_W,\,\sigma_{\infty}\}$ are in one-to-one correspondence to a unique holomorphic map $\beta$ from $\mathbb{T}^2$ to $\mathbb{P}^1$ marked with three points, say $\{0,1,\infty\}$.
This map is of degree $d$ and is ramified over the three points, with the ramification structure given by the permutations, associated by the natural identification $0 \leftrightarrow \sigma_B$, $1\leftrightarrow \sigma_W$, and $\infty \leftrightarrow \sigma_{\infty}$.

In this way, a cycle of length $n$ in a permutation corresponds to a point on $\mathbb{T}^2$ where the map is $n$-fold ramified.
If the cycle originates in $\sigma_B$ or $\sigma_W$, the length of the cycle is the ramification index, which is also the number of edges that extrude from the associated black or white node.
If the cycle originates in $\sigma_\infty$, the length of the cycle and the ramification index correspond to one-half the number of edges that surround the associated face of the dimer.
The meaning of the ramification index itself is quite simple.
In terms of local coordinates $w$ on the marked $\mathbb{P}^1$, which is the target of $\beta$, and local coordinates $z$ on $\mathbb{T}^2$, which one can think of as the source worldsheet, the map locally behaves as $w=z^n$, where $n > 1$ is an integer.
In turn, any continuous, non-self-intersecting segment on $\mathbb{P}^1$ connecting $0$ and $1$ with trivial monodromy around the point at $\infty$ is the image of the edges connecting the nodes in the dimer on $\mathbb{T}^2$.
Maps to $\mathbb{P}^1$ ramified over only $\{0,1,\infty\}$ are called \textbf{Belyi maps}, and the pair of data $(\mathbb{T}^2, \beta)$ is the \textbf{Belyi pair}.
The lesson to take home is that a dimer model (brane tiling),
a permutation triple and a Belyi pair are equivalent ways of completely capturing the information of a toric quiver gauge theory.

\paragraph{Example of ${\cal N}=4$ SYM:}
In order to make our discussion less abstract, let us illustrate the above concepts with the prototypical example of ${\cal N}=4$ super-Yang--Mills (SYM) theory, which arises as the worldvolume theory of D$3$-brane transverse to the trivial non-compact toric Calabi--Yau threefold $\mathbb{C}^3$.
This example is fully elaborated in~\cite{Jejjala:2010vb}.

\begin{figure}[h!]
\begin{center}
\includegraphics{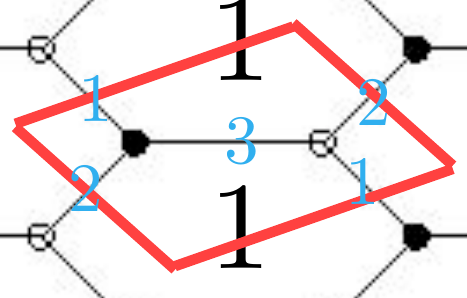}
\end{center}
\caption{
{\sf {\small Dimer for the $\mathcal{N}=4$ super-Yang--Mills theory, corresponding to the toric Calabi--Yau threefold $\mathbb{C}^3$.
There is only one gauge group $U(N)$, hence the single face, which is a hexagon marked by $1$.
There are three fields, all adjoints under this group, which emanate from the trivalent black/white nodes, labeled as $1$, $2$, and $3$.
The superpotential in terms of these three fields $\phi_{1,2,3}$ is the standard $W = {\rm Tr}(\phi_1\phi_2\phi_3 - \phi_1\phi_3\phi_2)$, as can be seen going around the white node clockwise and the black node anticlockwise.
The diagram is understood to extend doubly periodically and we have drawn, in red, the fundamental region.
}}}
\label{C3dimer}
\end{figure}

To begin, the dimer, with the labeled edges, is given in Figure~\ref{C3dimer}.
There is a total of three fields, all adjoints in this case, and we are therefore dealing with permutations in $S_3$.
The associated permutations can be easily read from Figure~\ref{C3dimer}, which in this case are very simple:
\begin{equation}
\sigma_B=(1\,2\,3) ~, \qquad \sigma_W=(1\,2\,3) ~, \qquad \sigma_{\infty}=(1\,2\,3) ~.
\end{equation}
As emphasized previously, we do not reverse direction when treating black and white nodes and therefore, in our convention, we proceed anti-clockwise for both the black and white trivalent nodes, thereby giving $\sigma_B$ and $\sigma_W$ as above.
The permutation cycle at infinity, $\sigma_\infty$ is obtained so that all three multiply to the identity in $S_3$.

The explicit expression for the Belyi pair is
\begin{equation}\label{belyiC^3}
y^2=x^3+1 ~, \qquad \beta=\frac{1+y}{2} ~,
\end{equation}
where the first is the $\mathbb{T}^2$ written as an elliptic curve in standard form, embedded in $\mathbb{C}[x,y]$ and the second is the Belyi map.
Following~\cite{Jejjala:2010vb}, it is straightforward to verify that this pair reproduces the combinatorial data encoded by the permutations.
\begin{equation}\label{c3eg}
\begin{array}{|c|c|c|c|c|c|}
\hline
\mathbb{T}^2: y^2 = x^3 + 1 & \stackrel{\beta=\frac12(1+y)}{\longrightarrow} & \mathbb{P}^1 & \mbox{Local Coordinates on }\mathbb{T}^2 & \mbox{Ramification Index of } \beta \\ \hline\hline
(0,-1) & \stackrel\beta\mapsto & 0 & (x,y) \sim (\epsilon,-1-\frac12\epsilon^3) & 3 \\ \hline
(0,1) & \stackrel\beta\mapsto & 1 & (x,y) \sim (\epsilon,1+\frac12\epsilon^3) & 3 \\ \hline
(\infty,\infty) & \stackrel\beta\mapsto & \infty & (x,y) \sim (\epsilon^{-2},\epsilon^{-3}) & 3 \\ \hline
\end{array}
\end{equation}

We can in fact make the correspondence even more explicit by looking at the pre-image under $\beta$ of the segment between $0$ and $1$ on $\mathbb{P}^1$.
Because the endpoint map to the black/white nodes by our construction, a simple non-self-intersecting curve connecting $0$ and $1$ with a trivial monodromy around $\infty$ should give precisely the edges in the dimer.
Let us consider the trivial curve $C(t) = t, t \in [0,\,1]$.
Then, the pre-image of such a segment is given by
\begin{equation}
y=2\,t-1 ~,
\end{equation}
on the elliptic curve $y^2 = x^3 + 1$.

In order to plot effectively,
let us resort to the standard
Weierstra\ss\ representation of an elliptic curve
in terms of the $\wp$-function.
We recall that the $\wp $ function gives the map between
the algebraic description of the torus and the description
as a quotient of the complex plane
modded out by a lattice (see, for example,
Theorem 6.14 of~\cite{Knapp}).
\begin{equation}\label{curve}
(x,y) = (\wp(z; \{g_2,g_3\}), \wp'(z; \{g_2,g_3\}))
\Longrightarrow y^2 = 4x^3 - g_2 x - g_3 ~,
\end{equation}
Here $g_2, g_3$ are the Weierstra\ss\ coefficients.

\begin{figure}[h!t!]
\begin{center}
\includegraphics[scale=.7]{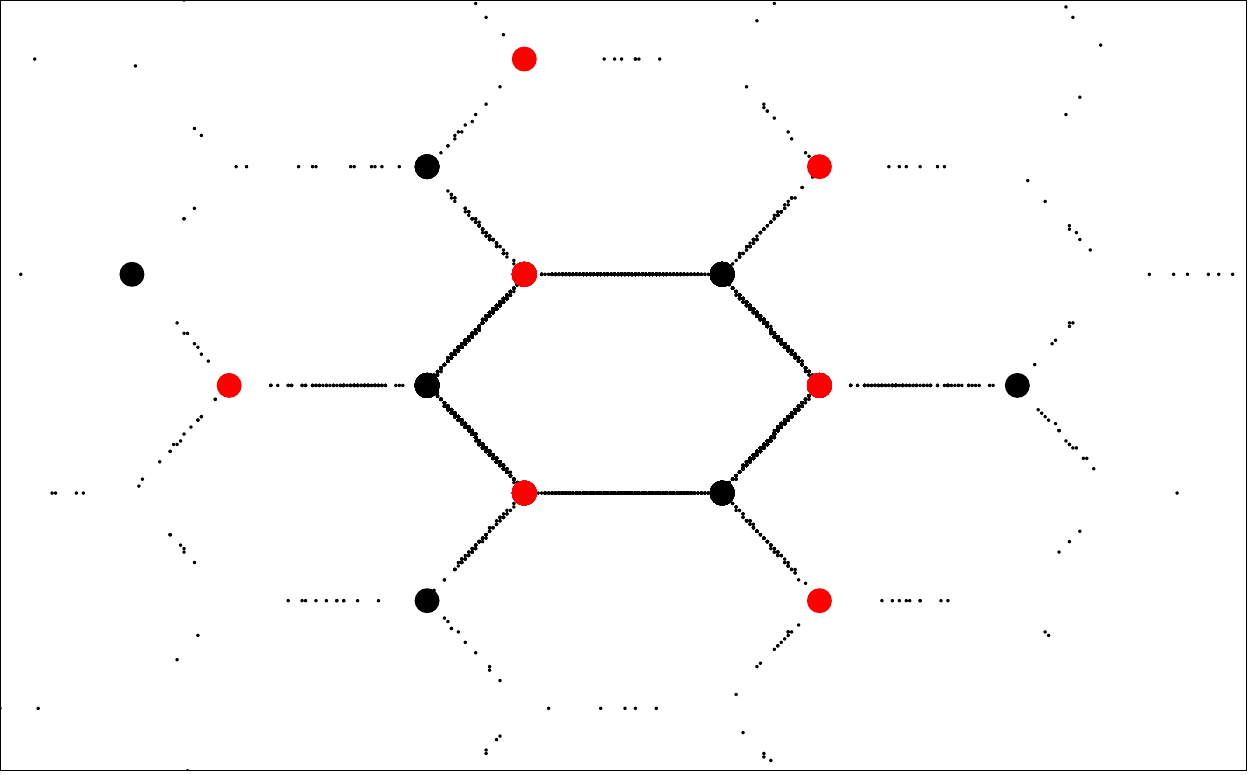}
\end{center}
\caption{{\sf {\small Pre-image of the interval $[0,\,1]$ in the $\mathbb{P}^1$ by the Belyi map $\beta$ explicitly recovers the dimer model in Figure~\ref{C3dimer}, a periodic honeycomb tiling of the plane for $\mathcal{N}=4$ super-Yang-Mills theory corresponding to $\mathbb{C}^3$.}
}}
\label{C3numeric}
\end{figure}

Indeed, for convenience, upon rescaling $(x,\,y)\mapsto (4\,x,\,4\,y)$, the Belyi curve for $\mathbb{C}^3$ becomes $y^2=4x^3+\frac{1}{16}$; that is, $\{g_2, g_3\} = \{0, -\frac{1}{16}\}$.
Thus, the explicit map from the fundamental domain of the $\mathbb{T}^2$ to the interval $[0,1]$ on the $\mathbb{P}^1$ is
\begin{equation}
4\,\wp'(z,\,\{0,-\frac{1}{16}\}) = 2\,t+1 ~, \qquad t \in [0,\,1] ~.
\end{equation}
By numerically solving this equation for each $t$, we can plot in the $z$-plane the pre-image of the interval $[0,\,1]$ which we show in Figure~\ref{C3numeric}.
This is, as expected, precisely the dimer model, which, as we recall from Figure~\ref{C3dimer}, lives in the fundamental region of the torus.

As in~\cite{Feng:2005gw}, a projection known as the alga map, complementing the amoeba map in tropical geometry, was constructed in order to obtain the dimer model explicitly from the mirror geometry to the toric threefold.
The above procedure of using the inverse of the Weierstra\ss\ $\wp$-function is an efficient method indeed of extracting the dimer from the
associated Belyi geometry.

\subsection{Isoradial dimers and the $\tau_R=\tau_B$ conjecture}

\begin{figure}[h!t!]
\begin{center}
\includegraphics[scale=.3]{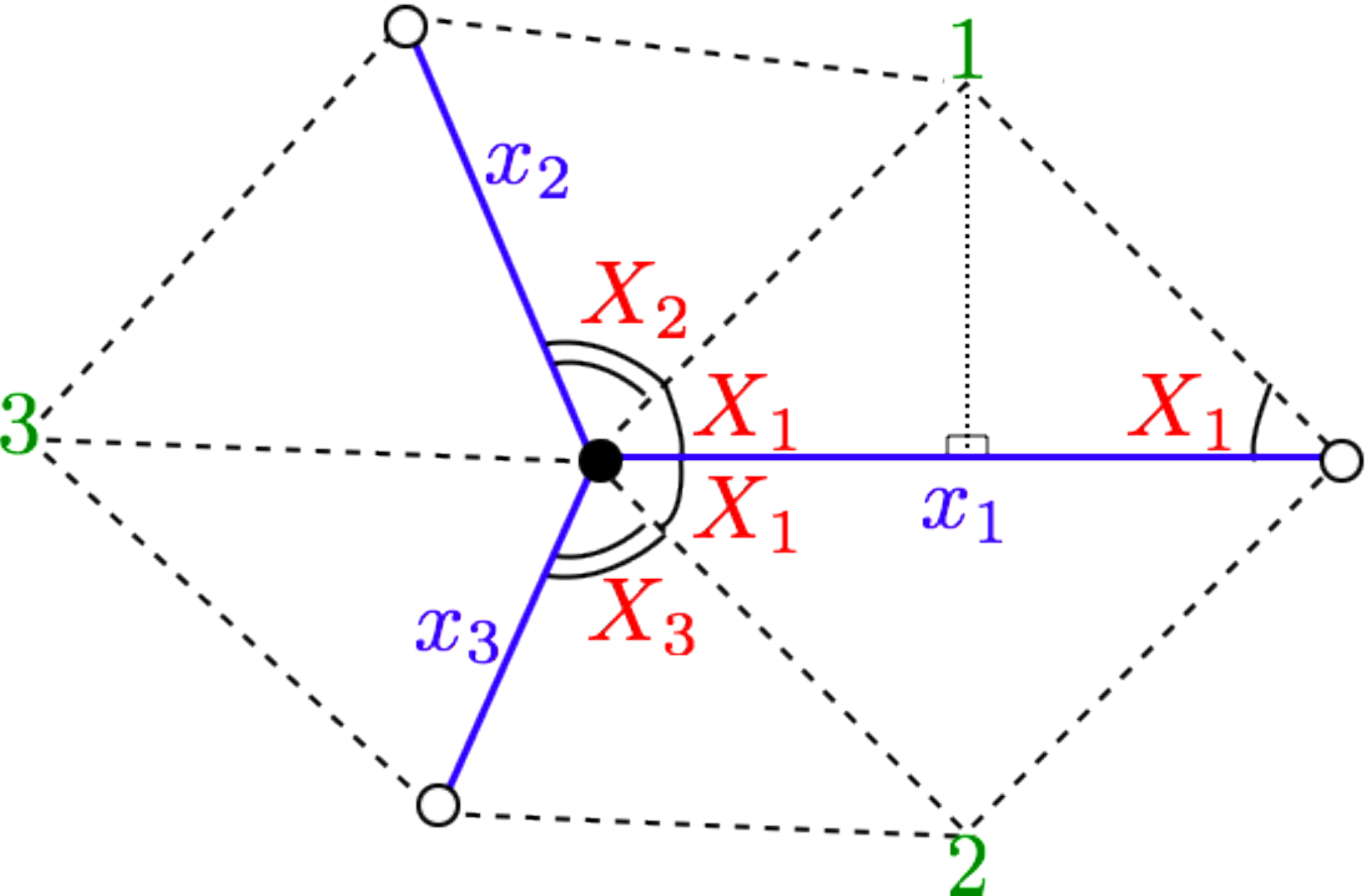}
\end{center}
\caption{{\sf \small Dimer in isoradial embedding. Blue lines are edges in the tiling. Green numbers, 1, 2, and 3, are centers of tiles 1, 2, and 3, respectively.
Dotted lines are all of equal length and stretched between centers of tiles to nodes of the dimer model.
Each dimer has an $R$-charge, \textit{e.g.}, the field corresponding to the edge $x_1$ has $R$-charge $R_1$,
then the angle $X_1$ is $\frac{\pi}{2}R_1$ and the angle from the center of tile 1 or tile 2 to the end points of $x_1$ is $\pi(1-R_1)$.
}}
\label{fig:angles}
\end{figure}

Before closing this lightning review of dimers and Belyi pairs, let us revisit the so-called \textbf{isoradial embedding} of the dimer.
In~\cite{Hanany:2005ss} the concept of isoradial embedding was introduced.
Following the mathematical literature (see, \textit{e.g.},~\cite{kenyon} for a review), it turns out to be highly useful to draw the dimer such that all nodes lie in circles of unit radii centered on the faces (hence the name isoradial). A fragment of such an embedding is shown in Figure~\ref{fig:angles}

There is a moduli space of isoradial embeddings.
As described in~\cite{Hanany:2005ss}, the angle subtended by an edge
of the face in an isoradially embedded dimer is $\pi(1-R)$
where $R$ is the $R$-charge of the field which corresponds to
that edge (see Figure~\ref{fig:angles}).
Thus, the fact that angles around a vertex sum to $2\pi$ and
that the sum of the angles subtended by the edges of a face sum
to $2\pi$ translate to the requirements that each monomial in the
 superpotential has $R$-charge two and that the $\beta$-function
 vanishes for each of the gauge groups.
In terms of the dimer models, this is the statement that we can
 draw the bipartite graph on a plane.
A distinguished element among the isoradial dimers
is determined by the $R$-charges of the fields in the SCFT.
which are obtained through the procedure of maximizing
the central charge $a$~\cite{Intriligator:2003jj}.)

We call the isoradial dimer fixed by the $R$-charges the \textbf{$R$-dimer}.
This graph lives on a particular torus whose modular parameter
 is called $\tau_R$.
Based on a number of examples and consistency checks, it
was conjectured in~\cite{Jejjala:2010vb} that the modular
parameter $\tau_R$ is $SL(2,\mathbb{Z})$ equivalent to the
modular parameter $ \tau_B$ of the elliptic curve $\mathbb{T}^2$ in the
Belyi pair. Although ${\cal N}=4$ SYM, the conifold, and their orbifolds
satisfy the $\tau_R=\tau_B$ conjecture, the equivalence turns out
to be false more generally as we shall demonstrate with an explicit
counterexample later in this note. However, the existence of a
 unique Belyi pair for each dimer, with its rich number theoretic
information, motivates a deeper study of analogous number theoretic
structures associated with the $R$-charges themselves, a theme to which
we return in~\cite{Seibergdualtaus}.

\section{Towards a general algorithm for constructing Belyi pairs}
\label{algorithm}\setall

As mentioned in the introduction, explicitly constructing a Belyi
pair corresponding to a bipartite graph on a genus $g$ Riemann surface
 is technically challenging, and no known method exists in general.
Indeed, a central problem in the program of studying
dessins proposed by Grothendieck~\cite{grotesquisse}, is to extract the
number theoretic properties contained in the explicit
 Belyi pair, such as fields of definition,
from the combinatoric data of the dessins.
Aided by modern computer algebra, we can at least algorithmically attack the problem for the case of genus one.
In this section, we will show how to do so in a systematic way;
many of the Belyi pairs we shall exhibit later will be constructed using the method outlined here.

First, let us choose a standard form for the elliptic curve $\mathbb{T}^2$:
\begin{equation}\label{Clam}
y^2 = x(x-1)(x-\lambda) ~, \qquad
j(\lambda) = \frac{256 (1 - \lambda + \lambda^2)^3}
{\lambda^2(1 - \lambda)^2} ~.
\end{equation}
For completeness, we have also written Klein's modular
invariant $j$-function in terms of the $\lambda$ parameter.
The above form emphasizes the values of $x$ where $y$ vanishes.
Another common standard (Weierstra\ss)
 form $ y^2 = 4 x^3 - g_2 x - g_3$ can be obtained by an elementary
 coordinate change~\cite{hartshorne}.

We see immediately that in the form~(\ref{Clam}), there are four distinguished points on the elliptic curve:
$(0,0)$, $(1,0)$, $(\lambda,0)$, and $(\infty, \infty)$.
Let us consider the finite special points $x_0\in \{0,1,\lambda\}$ where $y$ vanishes.
Near the points $(x_0, 0)$, our $\mathbb{T}^2$ will, under a linear perturbation, look like
\be
(x,y) = (x_0 + \delta x, \delta y) \Longrightarrow
(\delta y)^2 = (x_0+\delta x)(x_0 - 1 + \delta x)(x_0 - \lambda + \delta x) ~.
\ee
For any choice of $x_0$ the right-hand side will be linear in $\delta x$:
\begin{equation}
(\delta y)^2 = c\, \delta x ~,
\end{equation}
which implies that $\epsilon\equiv\delta y$ is a good local
coordinate: $(\delta x,\delta y) \sim (\epsilon^2, \epsilon)$.
Another special point is $(\infty, \infty)$. Near this point,
a good local coordinate $ \epsilon^{-1} $ gives
 $(x,y) \sim (\epsilon^{-2}, \epsilon^{-3})$.
In addition to these distinguished points we have
to be careful near the points where the first
 derivative of $x(x-1)(x-\lambda)$ vanishes.
 Locally these will look like:
\begin{equation}
y_0 + a\,\delta y = x_0 + b\,(\delta x)^2 ~,
\end{equation}
which means that we must pick $\epsilon\equiv\delta x$ as the good local coordinate: $(\delta x,\delta y) \sim (\epsilon, \epsilon^2)$.
Over any other point, which we shall call generic, either $x$
 or $y$ is a good local coordinate: $(\delta x,\delta y) \sim (\epsilon, \epsilon)$.

An important difference between the four distinguished points
$\{ (0,0), (1,0), (\lambda,0), (\infty, \infty) \} $ and other values of $x$
 is that these other values of $x$ give a pair of points $(x, \pm y)$
on the curve, whereas for the special values, there is only one
point on the curve for each $x$.

Next, let us adopt a convenient notation, inspired by the rightmost column of~(\ref{c3eg}), for encoding the ramification indices.
Let there be $W$ pre-images of $0$, $B$ pre-images of $1$ and $I$ pre-images of $\infty$;
thus in the dimer there will be $B$ black nodes, $W$ white nodes and $I$ polygonal faces in the fundamental domain.
These equate to the number of cycles in the corresponding permutation.
Letting the ramifications of the pre-images of the three marked points be, respectively,
$\{r_0(1) , r_0(2) , \ldots , r_0(B)\}$,
$\{r_1(1) , r_1(2) , \ldots , r_1(W)\}$, and
$\{r_\infty(1) , r_\infty(2) , \ldots , r_\infty(I)\}$,
we must therefore adjust the rational function $\beta$ to satisfy these data.
In summary, the input data of our Belyi pair will be denoted by
\begin{equation}\label{B-input}
y^2 = x(x-1)(x-\lambda) ~, \qquad
\left\{\begin{array}{l}
r_0(1) , r_0(2) , \ldots , r_0(B) \\
r_1(1) , r_1(2) , \ldots , r_1(W) \\
r_\infty(1) , r_\infty(2) , \ldots , r_\infty(I)
\end{array}
\right\} ~.
\end{equation}
We emphasize the constraint that
$\sum_i r_0(i) = \sum_i r_1(i) = \sum_i r_\infty(i)$, which is the
degree of the map.

Now let us proceed to our construction. We will first
address the class of Belyi maps which depend only on the coordinate $x$.
This is not as limited as one might assume upon first glance:
indeed most of the maps constructed in~\cite{Jejjala:2010vb},
with some notable exceptions including $\mathbb{C}^3$,
belong to this category. We will soon unveil infinite families of examples.
We will thus address this case first before moving on to the general situation.

\subsection{Belyi is $x$-dependent only } \label{xonly}

Let $\beta(x) = P(x) / Q(x)$ where $P$ and $Q$ are polynomials in $x$.
Thus written, the pre-images of $0$ and $\infty$ are manifest.
We may also assume, without loss of generality, that the degree of $P$ exceeds that of $Q$ since, after all, the reciprocal of a Belyi map is also Belyi, serving merely to shuffle the image points $(0,1,\infty)$. We will explain
in Appendix~\ref{P1belyi} that the $x$-only ansatz amounts to constructing
$\mT^2 \rightarrow \mP^1 $ Belyi maps from $ \mP^1 \rightarrow \mP^1 $
Belyi maps.

Let the pre-images
 of $0$ have coordinates $x = \{ z_{1}, \ldots, z_{B'} \}$, which we can define to be distinct.
Because $\beta$ is assumed to have numerator of higher degree, $(\infty, \infty)$ is always mapped to $\infty$, and we take all the $z_i$ to be finite.
Correspondingly, the number of black nodes in our dessin/dimer is determined as follows:
each generic value $z_i \notin \{0,1,\lambda\}$ contributes two (since, as mentioned, there would be two (square root) values of $y$ on the elliptic curve),
and each distinguished point $z_i \in \{0,1,\lambda\}$, merely one (since $y$ would be zero only for these values).
Thus $B$ can be determined from $B'$ by summing with these contributions.

Similarly, let there be $I'$ pre-images of $\infty$.
Recalling that $x = \infty$ is always by assumption one of the pre-images, let us set the finite pre-images to have coordinates $x = \{ d_{1} , d_{2} , \ldots, d_{{I'-1}} \}$, corresponding to the points where $Q(x)$ vanishes.
Again, $I$ and $I'$ are related by having double contribution from generic points and single contributions from the distinguished points.

The zeros at $z_i$ and poles at $d_i$ immediately fix the factorization of the Belyi map to be
\begin{equation}\label{ansatz-x}
\beta(x) = \frac{P(x)}{Q(x)} =
\frac{A \prod\limits_{i=1}^{B'}( x - z_i)^{m_i}}
{\prod\limits_{i=1}^{I'-1} (x - d_i)^{n_i}} ~,
\end{equation}
where $A$ is some overall complex number.
For finite values $z_i \notin \{0, 1, \lambda\}$,
where $ \delta x $ is a good local coordinate as discussed above,
the exponents $m_i$ are equal to $r_0(i)$.
For distinguished points $z_i = 0, 1$ or $\lambda$, the good local
coordinate is $ \delta y = (\delta x)^2 $,
so we need to divide by two to obtain $m_i = r_0(i)/2$.
Similarly, $n_i = r_\infty(i)$ for generic $d_i$ and $n_i = r_\infty(i)/2$
 for distinguished $d_i$.

Instantly, we encounter a situation that the ansatz fails to address.
Because any distinguished point must contribute an even power but any generic point will contribute two factors to the product, any odd ramification index without a partner will not be taken care of by the form~(\ref{ansatz-x}) and thus cannot depend on $x$ alone.
We will call such circumstances as having unpaired odd ramifications.
For our prototypical example of $\mathbb{C}^3$, the ramification structure, in our notation, is $\left\{ \begin{array}{l} 3\\3\\3 \end{array}\right\}$, which is clearly odd, unpaired for all three rows.
We see indeed that here the Belyi map depends on $y$.
A ramification structure of, for example,
$\left\{ \begin{array}{l} 3,3,4\\3,3,4\\2,2,3,3 \end{array}\right\}$
 is acceptable;
this is an example which we will encounter later.

With the form quite explicit and the pre-images of $0$ and $\infty$
 taken care of, we must ensure that the pre-images of $1$ are in accord
 with the data $\{r_1(1) , r_1(2) , \ldots , r_1(W)\}$.
One way of doing so is to take the derivative of $\beta(x)$ with respect
 to $x$ and make sure that the roots of $\beta'(x)$ vanish at a set of
 points, different from $z_i$ and $d_i$, such that the order of vanishing upon
 them is in accord with the ramification indices $r_1(i)$; these will
 constitute the appropriate pre-images $o_i$ of $1$. In order to achieve
 that on top of vanishing of $\beta'(o_i)$ we must also impose $\beta(o_i)=1$.
In all, we have all the positions of the $x$-coordinates $z_i,d_i,o_i$
as well as the constant $A$ to tune in order to find the Belyi map.
With this input data, we can search for Belyi maps using
a program such as {\tt Mathematica}.
We emphasize that the input data~(\ref{B-input}) do not fully
specify the Belyi pair and that only with the knowledge of the
 permutation cycles can the uniqueness theorem of Belyi apply.
As we shall see later, there are significantly different dimer
 models which share the ramification structure.

\begin{figure}[h!t!]
\begin{center}
\includegraphics[scale=.5]{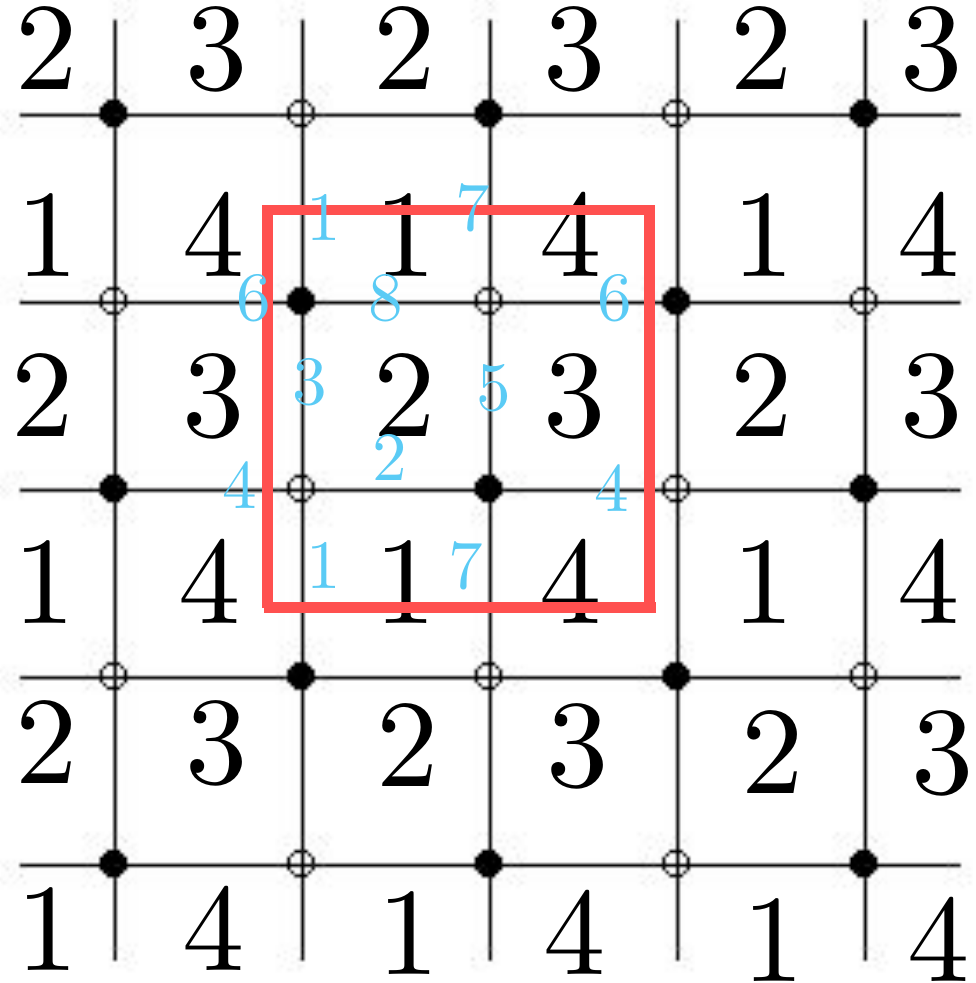}
\end{center}
\caption{{\sf \small Dimer for the phase 1 of $\mathbb{F}_0$.
}}
\label{fig:f0v1}
\end{figure}

Let us descend from the above abstraction with the illustration of
 a concrete example.
The first phase of the theory for the Calabi--Yau cone over the zeroth
 Hirzebruch surface $\mathbb{F}_0 \simeq \mathbb{P}^1 \times \mathbb{P}^1$
is a famous theory. The dimer for this theory is shown in
Figure~\ref{fig:f0v1}.
The ramification structure is
$\left\{ \begin{array}{l} 4,4\\ 4,4\\ 2,2,2,2\end{array}\right\}$.
Therefore, we can try $\beta(x) = A \frac{(x-a)^4}{x (x-1) (x-\lambda)}$
 for $a,b \notin \{0,1,\lambda\}$.
We have put the points $\{0,1,\lambda\}$ as zeros of the denominator by
 convenience since they provide three good points for infinity;
this forces, naturally, that the numerator does not have such factors.
A single generic factor $(x-a)^4$ suffices in the numerator as $x=a$ corresponds to two points on the elliptic curve.

Next, we need to solve for the critical points where $\beta'(x)=0$, demanded by
\begin{equation}
\partial_x \beta(x) =
-\frac{A (a-x)^3 \left(a \left(-2 (\lambda+1) x+\lambda+3 x^2\right)+x \left(-2 (\lambda+1) x+3 \lambda+x^2\right)\right)}{(x-1)^2 x^2 (\lambda-x)^2} = 0 ~.
\end{equation}
Clearly, $x=a$ would give a triple-critical points at which $\beta$ itself has image $0$.
We also need to make sure all other critical points map to $1$.
One immediate way is to enforce that the second factor
$\left(a \left(-2 (\lambda+1) x+\lambda+3x^2\right)+x \left(-2 (\lambda+1) x+3\lambda+x^2\right)\right)$
is a perfect cubic.
This happens, as one readily finds, when
$(a,\lambda) = (\pm i, -1)$,
$(\frac12 \pm \frac{i}{2}, \frac12)$, or
$(1 \pm i, 2)$.

We note that all the candidate values for $\lambda$ yield the same $j$-invariant, and thus they are all solutions on the same elliptic curve.
In fact, these various solution are equivalent to each other by redefinitions.
The solution $(-i, -1)$ is particularly eye-catching, since this would give us the Belyi pair
\begin{equation}\label{B-F0}
y^2 = x(x^2-1) ~, \qquad \beta(x) = \frac{i(i+x)^4}{8x(1-x^2)} \ ,
\end{equation}
exactly the one given in Section C.2.3 of~\cite{Jejjala:2010vb}.
Thus effortlessly we can generate algorithmically what once had to involve clever guesswork.

\subsection{The general problem}

It is tempting, given the essentially algebraic nature of our problem, to harness the power of computational algebraic geometry and computer algebra for large polynomial systems, and to develop a general method.
Though in principle we can do so, as we now show, the calculations involved quickly exceeds current computer capabilities.

Nevertheless, a strategy is clear.
We again start with~(\ref{B-input}), but now assume a form $\beta(x,y) = \frac{P(x) + Q(x) y}{S(x) + T(x) y}$, with $P,R,S,T$ some polynomials in $x$ yet to be fixed.
This is the most general form for the rational function $\beta$ because we recall that the equation of the elliptic curve will substitute any power of $y$ exceeding and including the quadratic in terms of successive cubics in $x$.
In fact, we can do better by multiplying the numerator and denominator by $S(x) - T(x) y$ so that the denominator becomes $S(x)^2 - T(x)^2 y^2$, which in turn is a function of $x$ only, by substituting the $y^2$ factor via the equation of the curve.
In summary, our ansatz for the Belyi pair will take the form
\begin{equation}
(\mathbb{T}^2, \beta(x,y)) = \left( y^2 = x(x-1)(x-\lambda) ~, \frac{P(x) + R(x) y}{Q(x)} \right) ~,
\end{equation}
where $P(x), Q(x), R(x)$ are polynomials in $x$ of sufficient degree so as to allow enough coefficients to adjust to the constraints.
It is nice to see that the denominator affords a form which can be factorized by the rules of~(\ref{ansatz-x}).

Next, it is expedient to introduce the \textit{total derivative}, which is the derivative to be henceforth used when considering the order of vanishing (\textit{i.e.}, ramification) at the branch points when restricted to $\mathbb{T}^2$.
Defining $F(x, y) = y^2 - x (x-1) (x-\lambda)$, which must vanish identically on the curve, we have that
\begin{equation}
\frac{d}{dx} = \frac{\partial}{\partial x} - \frac{\partial_x F}{\partial_y F} \frac{\partial}{\partial y} ~.
\end{equation}
This expression is valid at the points where $x$ is a good local coordinate. As noted before, this is not going to be the case at $x_0=0,1,\lambda$, which is reflected in the fact that $\partial_y F=2y$ vanishes at these points and the second term diverges. Therefore, alternatively, we can use:
\begin{equation}
\frac{d}{dy} = \frac{\partial}{\partial y} - \frac{\partial_y F}{\partial_x F} \frac{\partial}{\partial x} ~,
\end{equation}
which is valid when $\partial_x F \neq 0$ and thus $y$ is a good local coordinate. Finally, near the point $(\infty,\infty)$, where a good coordinate is $\epsilon$ with $x=1/\epsilon^2$ and $y=1/\epsilon^3$, the total derivative can be written as
\begin{align}
	\frac{d}{d\epsilon} =
	-2 y \frac{\partial}{\partial x}
	-3 x^2 \frac{\partial}{\partial y} ~.
\end{align}
If $\beta(\infty)=\infty$, which is the case in our constructions, then this derivative is understood to be acting on $1/\beta$, which is a good local coordinate in the target space.

Now that we have to specify both $(x,y)$ we no longer have the issue with the doubling of factors mentioned in the case when $\beta$ is a function of $x$ only.
Therefore, we need only adhere to a straightforward routine as follows.
First, let $(x_0^i, y_0^i)$ be a pre-image of $0$ with ramification $r_0(i)$, this means that
$\left.\frac{d^k}{dx^k} \right|_{(x_0^i, y_0^i)} \beta(x,y) = 0$ for all $k = 0,1,2,\ldots, r_0(i) - 1$, where $k=0$ is just evaluation.
This gives us $r_0(i)$ algebraic conditions, in addition to the condition that $(x_0^i, y_0^i)$ needs to reside on $\mathbb{T}^2$.
We must do this for each of the ramification points at $0$, and then at $1$, for which the zeroth-order term should be set to $1$.
Finally, the denominator can have product form over where the map is allowed to blow up.
The above was under the assumption that the pre-images are generic points.
We could also repeat this, for all combinations, whenever we have an even ramification index, by demanding the vanishing of powers of the total derivative up to half of the ramification over distinguished points.

We see, of course, that this algorithm will rapidly produce significant number of polynomials in all the pre-image points as well as the undetermined coefficients in $P(x)$, $Q(x)$, and $R(x)$.
Since reduction of such polynomials systems of non-trivial degree requires exponential running time, even with the entire subject of computational algebraic geometry and commutative algebra at disposal, the systems will quickly defy analysis and render our brute force approach impractical.
We will soon see how to utilize the power of orbifolding techniques to construct many Belyi pairs, bypassing, therefore, this computational hurdle.

On a more optimistic note to conclude this section of the construction of Belyi pairs, we can use this aforementioned algorithm to reproduce the $\mathbb{C}^3$ result instantly.
Starting with the ansatz of $P$, $R$ being linear in $x$, say, and $Q$ being a constant suffices (recalling that $(\infty,\infty)$ is here mapped to $\infty$, so there need not be any finite factors in the denominator).
We can solve for the value of $\lambda$ to be $\frac12(1+i \sqrt{3})$ and
$\beta = \frac12 (1+(-1)^{1/4}3^{3/4}y)$.
It is easy to see that upon the simple coordinate transformation $x' = \frac12(3+i \sqrt{3}) x - 1; \ y' = (-1)^{1/4} 3^{3/4}y$ brings the Belyi pair to the requisite form introduced in~(\ref{belyiC^3}).

\section{Orbifolds and covering the torus}
\label{orbifolds}\setall

A natural operation in string theory and in algebraic geometry is orbifolding, the identification of points by the action of a finite group on a given geometrical space.
When D$3$-branes probe an orbifolded geometry, the finite group acts on both the boundary conditions and the Chan--Paton factors of the open strings, thus giving rise to a ``daughter'' orbifolded theory in the fashion described in~\cite{Douglas:1996sw, Kachru:1998ys, Lawrence:1998ja}.

From the dimer perspective, orbifolding is represented in a very simple manner.
As described in~\cite{Hanany:2005ve}, starting from the periodic tiling corresponding to a given dimer describing the SCFT for a Calabi--Yau space, the operation of orbifolding the Calabi--Yau amounts to an enlargement of the unit cell of the tiling.
This corresponds to going to an unbranched
 cover of the $\mathbb{T}^2$, which in turn has implications for the associated Belyi pairs.

\subsection{Unbranched covers of the torus}

We recall that the Belyi map is a branched cover from $\mathbb{T}^2$ to $\mathbb{P}^1$:
\begin{equation}
\beta:\ \mathbb{T}^2\,\rightarrow \, \mathbb{P}^1 ~,
\end{equation}
with degree $d$ equal to the number of edges in the dimer and ramifications over $\{ 0, 1, \infty\} $ related to the structure of the dimer.
Consider now an unbranched cover $\widehat{\mathbb{T}}^2$ of the elliptic curve $\mathbb{T}^2$:
\begin{equation}
\psi:\ \widehat{\mathbb{T}}^2\, \rightarrow \mathbb{T}^2 ~,
\end{equation}
of degree $n$.
Such a map $\psi$ has $n$ inverse images for every point on $\mathbb{T}^2$.
Its derivative is nowhere vanishing, which is what we mean when we write that the cover is unbranched.
These properties are clear from the picture of enlarging the unit cell.

\begin{figure}[h]
\centering
\includegraphics[scale=0.4]{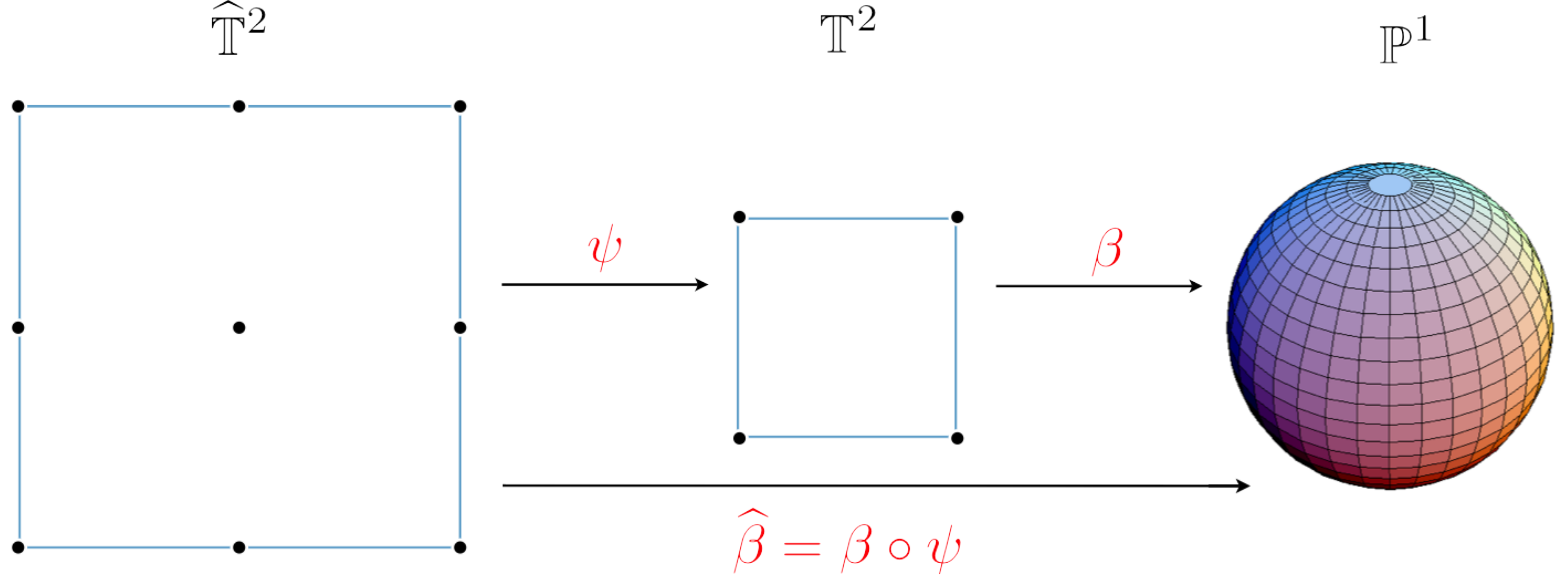}
\caption{{\sf \small The composition of an unbranched cover of the torus and a Belyi map is Belyi.
}}
\label{widehatbeta}
\end{figure}

The composition of $\beta$ and $\psi$
\begin{equation}
\widehat\beta: \widehat{\mathbb{T}}^2 \rightarrow \mathbb{P}^1 ~,
\qquad
\mbox{ where }
\widehat\beta=\beta\circ \psi~,
\end{equation}
is interesting, as we now shall see.
The operation is indicated in Figure~\ref{widehatbeta}.
With a local coordinate $z$ on $\widehat{\mathbb{T}}^2$, we have that
\begin{equation}
\partial_z \widehat\beta = \partial_z \beta ( \psi ( z )) = \partial_\psi \beta\ \partial_z \psi ~,
\end{equation}
by application of the chain rule.
Because the cover is unbranched, $\partial_z \psi \ne 0$.
The only zeros of $\partial_z \widehat\beta$ therefore occur when $\partial_\psi \beta = 0$.
Since $\beta$ is Belyi, its derivative only vanishes at points where $\beta ( \psi ) \in \{ 0 , 1, \infty \}$ and so $\widehat\beta(z) \in \{ 0, 1, \infty \}$ whenever $\partial_z \widehat\beta$ vanishes.
Thus, $\widehat\beta$ is also a Belyi map.
Consequently, we have constructed a new Belyi pair $(\widehat{\mathbb{T}}^2, \widehat\beta)$, and there should be a corresponding dimer model.

Each ramification point of $\beta$ lifts to $n$ ramification points of $\widehat\beta$ with the same ramification index.
The number of faces of the new dimer is $n$ times that of the original dimer.
This translates into multiplying the number of factors in the gauge group by $n$ as expected~\cite{Douglas:1996sw}.
Thus, $(\widehat{\mathbb{T}}^2 , \widehat\beta)$ is the Belyi pair associated to the orbifolded SCFT.

\paragraph{Complex structure of the cover and the $\tau_R=\tau_B$ conjecture:}

\begin{figure}
\centering
\includegraphics[scale=1]{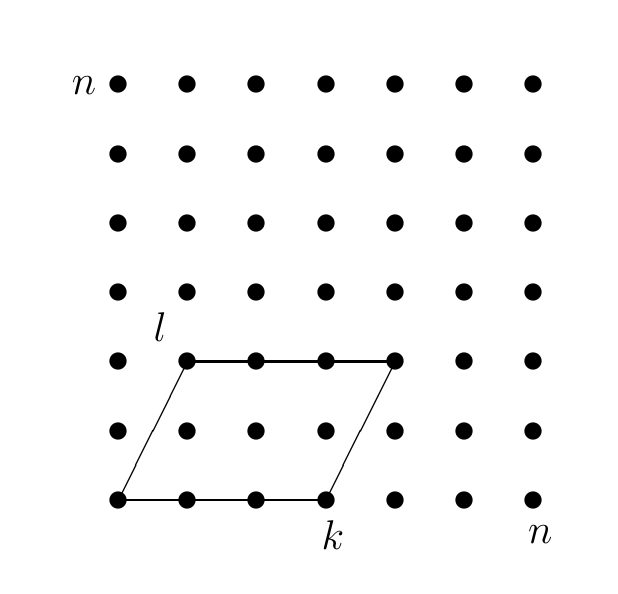}
\caption{{\sf \small The unit cell for a torus $\mathbb{T}^2$ is a drawn as a primitive lattice square.
The $n$-fold unbranched torus $\widehat{\mathbb{T}}^2$ can have its unit cell being any of the lattice parallelograms marked by $(p,l,k)$ with $pk=n$.
}}
\label{toruscover}
\end{figure}

The complex structure of the cover $\widehat{\mathbb{T}}^2$ can be described as a function of the complex structure of $\mathbb{T}^2$.
The covers of degree $n$ are known to be in one-to-one correspondence with
 integers $k , p, l$, such that $k \, p = n $, $ k , p > 0 $ and
 $ 0 \le l \le k -1 $.
This is a fact from the mathematics literature, and we refer the reader to, \textit{e.g.},~\cite{Gross:1993yt}).
These integers are indicated in Figure~\ref{toruscover}, with
$p$ being the height of the parallelogram.
For each such $(p,k,l)$, the complex structure $\tau_{\rm cover}$ of the cover $\widehat{\mathbb{T}}^2$ is given
in terms of the complex structure $\tau$ of the target torus $\mathbb{T}^2$ by
\begin{equation}
\label{tauofcover}
\tau_{\rm cover} (p,k,l) = \frac{l + p \, \tau}{k} ~.
\end{equation}

We note that, by construction, if the parent theory
satisfies $\tau_R=\tau_B$, so, too, will the orbifolded daughter theories.
The enlargement of the unit cell determines
the unit cell for the orbifold, hence $ \tau_R $.
The cell enlargement also determines the complex structure
for the covering torus, which is described algebraically by a map $ \psi $,
used in the construction of the Belyi map $ \hat \beta $ for the orbifold
theory.

\subsection{Constructing unbranched covers of tori}
\label{chicha}

As we have seen, the orbifolding of a theory reduces to the construction of $n$-fold unbranched covers of the torus where the parent theory Belyi map lives.
In order to explicitly construct those covers, let us consider a $\mathbb{T}^2$ defined by an elliptic curve $K$ which we will write in Weierstra\ss\ form.

\comment{ \footnote{
For readers who prefer the Tate form
\[
y^2+a\,x\,y+b\,y=c\, x^3+d\,x^2+e\,x+f
\]
of the elliptic curve, by performing the following change of variables
\[
(x,\,y)\ \mapsto\ (\frac{x}{c}-\frac{a^2+4\, d}{12\,c},\,y-\frac{a\,x}{2}-\frac{b}{2})
\]
we can bring it to the Weierstra\ss\ form.
}as we have done in~(\ref{curve}).
}

The Weierstra\ss\ elliptic function $\wp(z; \{g_2 , g_3\})$
can be used to map the description of the torus as a quotient
$ \mC / ( \mZ (2\omega_1) + \mZ (2\omega_2) ) $ by the lattice
generated by $ ( 2 \omega_1 , 2 \omega_2 ) $. The coefficients
$g_2,g_3$ are functions of these half-periods $ \omega_1 , \omega_2$,
which can be made explicit by writing $ g_2 ( \omega_1 , \omega_2 ) , g_3 ( \omega_1 , \omega_2 ) $. The pair
\begin{equation}
X = \wp(z; g_2, g_3) ~, \qquad
Y = \wp'(z; g_2, g_3 ) ~,
\end{equation}
where the prime denotes a derivative with respect to $z$,
obey the equation $ Y^2 = 4 X^3 - g_2 X - g_3 $.
The periodicities are
\bea
 \wp(z; g_2 ( \omega_1 ,\omega_2 ) , g_3 ( \omega_1 , \omega_3 ) ) & = &
\wp(z+ 2 \omega_1 ;
 g_2 ( \omega_1 ,\omega_2 ) , g_3 ( \omega_1 , \omega_3 ) ) \cr
& = &
\wp(z+ 2 \omega_2 ; g_2 ( \omega_1 ,\omega_2 ) ,
g_3 ( \omega_1 , \omega_3 ) ) ~, \nn \\
\eea
with the same periodicities holding for $\wp'(z; g_2, g_3 )$.
Any even meromorphic function of $z$ with the periodicities
$ ( 2 \omega_1 , 2 \omega_2 ) $ can be written as
a rational function
$ P ( X ) \over Q (X ) $.
Any odd meromorphic function of $z$ can be written as
$ R ( X ) Y \over S ( X ) $.

Let us consider now unbranched $n$-covers of the torus.
From the point of view of the $z$-plane, we are considering
for a given lattice, sublattices where the unit cell
has $n$-times larger than that of the original lattice.
From the cubic Weierstra\ss\ equation we are looking
for the equation of the covering curve
\bea
Y^2 = 4 X^3 - G_2 X - G_3
\eea
given
\bea
y^2 = 4 x^3 - g_2 x - g_3
\eea
and the covering map gives $ ( x , y ) $ as a function of
$( X , Y ) $.
 The Weierstra\ss\ function $x$ can be written
in terms of $X$ by summing the $X$ at fractional arguments
of the periods so as to produce the refined periodicity
(see, for example,~\cite{duval, wolframfuns}).
 It follows that the point $x = \infty $ maps to the point $ X = \infty $.
Using the general result that rational functions of
 Weierstra\ss\ functions $X$ gives rise to all even meromorphic functions,
we are led to consider
\begin{equation}\label{x-trans}
x (z) = \frac{P_n(X(z) )}{Q_{n-1}(X(z) )} ~,
\end{equation}
with $P_n$ and $Q_{n-1}$ being two arbitrary monic (top coefficient of the term $X^n$ and $X^{n-1}$ is one) polynomials of degree $n$ and $n-1$, respectively, in the variable $X$.
This ensures that $x = \infty$ maps to $ X = \infty $.
Regarding the elliptic curve as an abelian group, the Weierstra\ss\
function also maps the group operation to the addition of
the $z$-argument, so that addition theorems for Weierstra\ss\
functions give the group law. The point at infinity is
the identity element since the Weierstra\ss\ function has a double pole
at $ z = 0 $.

\comment{
Again, by doing $(x,\,y)\mapsto (4\,\tilde{x},\,4\,\tilde{y})$ the curve goes into
\begin{equation}
\label{curveP}
\tilde{y}^2=4\,\tilde{x}^3+\frac{a_4}{4}\,\tilde{x}+\frac{a_6}{16} ~.
\end{equation}
In these coordinates it is possible to solve the curve equation by means of the Weierstra\ss\ elliptic functions.
\begin{equation}
\tilde{x}=\wp(z;\,g_1,\,g_2) ~, \qquad \tilde{y}=\wp'(z;\,g_1,\,g_2) ~,
\end{equation}
where the prime denotes a derivative with respect to $z$ and $(g_1,\,g_2)$ are to be tuned depending on the constants $(a_4,\,a_6)$.
In the following we will omit such dependence.
Coming back to the original coordinates, we can write
}

The transformation on $x$,~(\ref{x-trans}), yields a transformation on $y$:
\be
y(z)=x'(z) =
 \frac{d}{dz} \frac{P_n(X(z))}{Q_{n-1}(X(z))}=\frac{\dot{P}_n\,Q_{n-1}-\dot{Q}_{n-1}\,P_n}{Q_{n-1}^2}\, X'(z) ~,
\end{equation}
where the dot represents $x$-derivative and the prime, the $z$-derivative.
Therefore, dropping the $z$-dependence, we have
\begin{equation}
y = \frac{\dot{P}_n\,Q_{n-1}-\dot{Q}_{n-1}\,P_n}{Q_{n-1}^2}\, Y ~.
\label{y-trans}
\end{equation}
\comment{
As a check, if we look at the point at infinity $P_n\sim x^n$ while $Q_{n-1}\sim x^{n-1}$, so
\begin{equation}\label{y-trans}
y\ \mapsto\ \frac{n\,x^{2\,(n-1)}-(n-1)\,x^{2\,(n-1)}}{x^{2\,(n-1)}}\, y\sim y
\end{equation}
and as desired since the point at infinity is mapped to itself.}

We now have a complete description of the map
 from an $n$-fold cover to the torus,
 in terms of the affine coordinates in Weierstra\ss\ form, namely equations~(\ref{x-trans}) and~(\ref{y-trans}).
Substituting these into the original curve in~(\ref{curve}) and simplifying, we find that
\begin{equation}
Y^2=\frac{4 P_n^3\, Q_{n-1}-g_2\,P_n\,Q_{n-1}^3-g_3\, Q_{n-1}^4}{\Big(\dot{P}_n\,Q_{n-1}-\dot{Q}_{n-1}\,P_n\Big)^2} ~,
\end{equation}
as the affine equation of the unbranched $n$-fold covering torus.
Imposing the Weierstra\ss\ form
\begin{equation}
\frac{4 P_n^3\, Q_{n-1}-g_2\,P_n\,Q_{n-1}^3-g_3\, Q_{n-1}^4}{\Big(\dot{P}_n\,Q_{n-1}-\dot{Q}_{n-1}\,P_n\Big)^2}=4X^3-G_2\,X -G_3
\end{equation}
for some new coefficients $(G_2,\,G_3)$.
\comment{
Note that one might wonder why we considered the RHS of this expression to be include generically an $x^2$ term, \textit{i.e.} $x^3+A_2\,x^2+A_4\,x+A_6$. However, the transformation removing this term is just a shift in $x$. Thus, with no loss of generality, we can just assume we consider this appropriate coordinate where $A_2=0$.
}
In other words, we need to solve identically in $x$, that
\begin{equation}
\label{mastereq}
4 P_n^3\, Q_{n-1}-g_2\,P_n\,Q_{n-1}^3-g_3\, Q_{n-1}^4-\Big(4X^3-G_2\,X-G_3\Big)\, \Big(\dot{P}_n\,Q_{n-1}-\dot{Q}_{n-1}\,P_n\Big)^2=0 ~.
\end{equation}

The left-hand side is a polynomial in $X$ of degree $4n-1$.
Since the above equation must hold for arbitrary $X$,
 it must be that the $4n$ coefficients of the left-hand side vanish.
On the other hand, the meromorphic transformation of $(X,\, Y)$ was expressed in terms of the $P_n,\, Q_{n-1}$ polynomials.
Also, as the coefficient of the highest power is one, these polynomials involve $n$ and $n-1$ \textit{a priori} unknown constants, respectively.
Finally, since we also need to fix $(G_2,\,G_3)$,
 in fact our transformation involves $2n+1$ unknowns.
The $4n$ equations form an over-complete system. However, we are guaranteed
the existence of multiple solutions since we know that the counting of
inequivalent coverings of tori, which are given by integers $[ k , l, p ] $
obeying $ kl = n ; k , l > 0, 0 \le l \le k-1 $
~\cite{Hanany:2010cx, Davey:2010px, Hanany:2010ne, newpaper}. The precise matching
of the solutions of the polynomial equation~(\ref{mastereq})
with the $[k,l,p]$ data is not trivial. It will be given in
examples for small $n$ in the following.

The above discussion in terms of polynomials $P_{n}( X ) , Q_{n-1} ( X ) $
is adequate for computations, and is similar in complexity to
our constructions for general Belyi maps from torus in Section~\ref{algorithm}. Further uses of the beautiful theory of Weierstra\ss\ functions
allow somewhat more explicit algorithms studied under the heading
of transformation theory of elliptic functions and modular relations, e.g
\cite{duval}. Most of the key formulae, such as those for
 fractional modifications of a period, integer multiplication of
the $z$ argument, are collected in~\cite{wolframfuns}. Nevertheless
completely explicit general formulae for the polynomials
at general $n$ matched with covering space data $ [ k , l , p ] $
remain elusive enough to have a role in cryptography
(see, for example,~\cite{bomosasch} for improved algorithms and for
references to the literature on applications). In the following,
we work out some examples using the direct method described above.
In an appendix, we describe some key equations
and applications from the elegant method of~\cite{duval}.

\subsection{Example: Degree $2$ covers of $y^2=x^3+1$}
\label{A1}

Consider the particular case of our familiar example $y^2=x^3+1$.
Following the above prescription, the transformation corresponding to degree two covers (at most quadratic in the numerator of~(\ref{x-trans})) must be
\begin{equation}
(x,\,y)\ =\ \left(\frac{X^2+\alpha_1\,X+\alpha_0}{X+\beta_0},\, \frac{X^2+2\,\beta_0\,X+(\alpha_1\,\beta_0-\alpha_0)}{(X+\beta_0)^2}\, Y\right) ~.
\end{equation}
Upon substituting this generic form in the curve equation and specializing to the case~(\ref{mastereq}), we find a number of different solutions in addition to the trivial one.

Solving for the coefficients, we see that there are three non-trivial transformations:
\begin{equation}
(x,\,y)\ =\ \left(\frac{3\,\omega^{2\,n}+8\,\omega^{-2\,n}\,X-16\,X^2}{8\,(\omega^{-2\,n}-2\,X)},\, \frac{X^2-X\,\omega^{-2\,n}+\frac{7}{16}\omega^{2\,n}}{(X-\frac{1}{2}\,\omega^{-2\,n})^2}\,Y \right) ~,
\end{equation}
with
\begin{equation}
\omega^3 = -1 ~, \quad n=\{-1,\,0,\,1\} ~.
\end{equation}
These give us the possible covering solutions
\begin{equation}
Y^2=X^3-\frac{15}{16}\,\omega^{2\,n}\,X+\frac{11}{32} ~.
\end{equation}
Using the standard formula,
\begin{equation}
j = 1728 \frac{g_2^3}{g_2^3 - 27 g_3^2} ~,
\end{equation}
we see that this curve has $j$-invariant 54000.
On the other hand, the original curve had vanishing $j$-invariant, which corresponds to $\tau=e^{i\frac{\pi}{3}}$.
For a degree two cover with $ [k , l , p ] = [1,2,0] $ from
 Figure~\ref{toruscover}, we expect $\tau$ to become $\tau_{\rm cover}=2\,\tau=2\,e^{i\frac{\pi}{3}}$.
Indeed, one can check that $j(\tau_{\rm cover})=54000$.
Since the Klein $j$-invariant classifies equivalence classes
 of elliptic curves, we have indeed arrived at the correct one.

\section{$\mathbb{C}^3$ and its orbifolds}
\label{C3example}\setall

Following the above discussion of
 constructing Belyi pairs and extracting $n$-fold covers, we now
 turn to concrete examples.
Again, let us begin with our familiar $\mathbb{C}^3$, whose Belyi pair we recall from equation~(\ref{belyiC^3}).
We construct various Abelian orbifolds by following the technology developed above.

\subsection{$\mathbb{Z}_2\times \mathbb{Z}_2$ orbifolds and period doubling}

Let us start with examining the $\mathbb{Z}_2\times \mathbb{Z}_2$ orbifold of $\mathbb{C}^3$.
Since the degree of the cover is four, we should consider the transformation
\begin{equation}
x\ =\ \frac{X^4+\alpha_3\,X^3+\alpha_2\,X^2+\alpha_1\,X+\alpha_0}{X^3+\beta_2\,X^2+\beta_1\,X+\beta_0} ~,
\end{equation}
as well as a corresponding expression for $y$.
Plugging these into~(\ref{mastereq}), we can find all the solutions.
After some algebra, we deduce that
\begin{equation}
\label{Z2Z2}
(x,\,y)\ =\ \left(\frac{8\,X\,(-1+8\,X^3)}{(1+64\,X^3)},\, \frac{8\,(-1+160\,X^3+512\,X^6)}{(1+64\,X^3)^2}\,Y \right) ~,
\end{equation}
with the resulting Belyi pair:
\begin{equation}
Y^2=X^3+\frac{1}{64} ~, \qquad
\beta=\frac{(Y+3)^3\,(Y-1)}{16\,Y^3} ~.
\end{equation}
One can easily check, has the correct combinatorial data for the expected $\mathbb{C}^3/\mathbb{Z}_2\times \mathbb{Z}_2$, namely by checking the order of vanishing at the pre-images of 0,1 and $\infty$, its ramification structure is precisely $\left\{ \begin{array}{l} 3, 3, 3, 3 \\ 3, 3, 3, 3\\ 3, 3, 3, 3 \end{array}\right\}$.

\subsubsection{Double angle and a consistency check}

We can cross check our result by noting
that the $\mathbb{Z}_2\times \mathbb{Z}_2$ orbifolding
 amounts to a doubling the unit cell in the two directions in the complex plane.
This procedure is a classical one known as \textit{doubling the period}, and our transformation must thereby be realized.
This is the reason why we have chosen the $\mathbb{Z}_2\times \mathbb{Z}_2$ orbifold as our initial example.
Recalling that our Weierstra\ss\ curve
\begin{equation}
\bar{y}^2=4\,\bar{x}^3-g_2\,\bar{x}-g_3
\end{equation}
has $(\bar{x}, \bar{y})=(\wp(z), \wp'(z))$.
Applying the period-doubling formula~\cite{wolframfuns} for the $\wp$-function:
\begin{equation}
\wp(2\,z)=\frac{\Big(\wp(z)^2+\frac{g_2}{4}\Big)^2+2\,g_3\,\wp(z)}{4\,\wp(z)^3-g_2\,\wp(z)-g_3} \ ;
\end{equation}
hence, in terms of the $x$, this is the transformation
\begin{equation}
\label{DA}
\bar{x}\ =\ \frac{\Big(\bar{X}^2+\frac{g_2}{4}\Big)^2+2\,g_3\,\bar{X}}{4\,\bar{X}^3-g_2\,\bar{X}-g_3} ~.
\end{equation}

To get rid of the coefficient $4$ in front of the $\bar{x}$, let us call $\bar{y}=\frac{y}{\sqrt{2}}$ and $\bar{x}=\frac{x}{2}$, so that the curve looks like
$y^2=x^3-g_2\,\bar{x}-2\,g_3$,
while the transformation is
$x\ =\ \frac{1}{4}\, \frac{(X^2+g_2)^2+16\,g_3\,X}{X^3-g_2\,X-2\,g_3}$.
Applying~(\ref{DA}) to our $\mathbb{C}^3$ example with $g_2=0$ and $g_3=1$ gives us
\begin{equation}
\label{deg4}
(x,\,y)\ =\ \left(\frac{X\,(X^3-8)}{4\,(X^3+1)},\,\frac{-8+20\,X^3+X^6}{8\,(X^3+1)^2}\,Y \right)
\end{equation}
It is straightforward to check that this leaves invariant the curve. It should then correspond to doubling the unit cell in the two directions, such that the $\tau$ remains unchanged.
Finally, upon rescaling $(X,\,Y)\mapsto (\frac{X}{4},\,\frac{Y}{8})$ in our transformation~(\ref{Z2Z2}), we recover exactly the map~(\ref{deg4}).
In other words, the four-fold cover is indeed consistent with the period doubling.

\subsection{Degree $3$ covers and $dP_0$}

Emboldened by our success, we can move on to further orbifolds.
The most famous one is undoubtedly the $\mathbb{Z}_3$ orbifold of $\mathbb{C}^3$, otherwise known as the Calabi--Yau cone $dP_0$ over the zeroth del Pezzo surface, which is simply the complex projective plane.
As the degree is three, we should consider
\begin{equation}
x\ =\ \frac{X^3+\alpha_2\,X^2+\alpha_1\,X+\alpha_0}{X^2+\beta_1\,X+\beta_0}
\end{equation}
along with the corresponding expression for $y$.
Clearly, there are various degree three covers, depending on how the unit cell is enlarged.
We would like however to find $dP_0$, which corresponds to a cover with the same $\tau$ as the original curve.

Thus, plugging the above transformation into~(\ref{mastereq}) and imposing that we want covers with the same $\tau$, we easily find
\begin{equation}
(x,\,y)\ =\ \left(\frac{X^3-4}{27\,X^2},\,\frac{27\,X^3+8}{27\,X^3}\,Y\right) ~,
\end{equation}
giving the curve
\begin{equation}
Y^2=X^3-\frac{1}{27} ~.
\end{equation}
It is easily checked that the rescaling $(X,\,Y)\mapsto (-\frac{X}{3},\,\frac{i}{3\,\sqrt{3}}\,Y)$ gives us the Belyi pair:
\begin{equation}
Y^2=X^3+1 \, \qquad
\beta = \frac{ \frac{i}{6\sqrt{3}} Y^3 +\frac{Y^2}{2} - \frac{i\sqrt{3}}{2} Y - \frac{1}{2} }{ (Y-1) (Y+1) } ~,
\end{equation}
which can be shown to yield to the correct ramification structure
$\left\{ \begin{array}{l} 3,3,3 \\ 3,3,3\\ 3, 3, 3\end{array}\right\}$.
Furthermore, this Belyi pair was previously constructed in~\cite{Jejjala:2010vb}, and we reproduce this result.

\subsection{Degree $2$ covers and $\mathbb{Z}_2$ orbifold}

Let us finally construct the cover corresponding to the $\mathbb{Z}_2$ orbifold of $\mathbb{C}^3$, or more precisely, $\mathbb{C}^2/\mathbb{Z}_2 \times \mathbb{C}$.
This is especially interesting, as the corresponding field theory is the $A_1$ $\mathcal{N}=2$ SCFT.
Applying the results from Section~\ref{A1}, the appropriate transformation is
\begin{equation}
(x,\,y)\ =\ \left(\frac{3+8\,X-16\,X^2}{8\,(1-2\,X)},\, \frac{X^2-X+\frac{7}{16}}{(X-\frac{1}{2})^2}\,Y \right) ~,
\end{equation}
and the resulting curve is
$Y^2=X^3-\frac{15}{16}X+\frac{11}{32}$.

Substituting the transformation rule into the Belyi map for $\mathbb{C}^3$ yields the Belyi pair:
\begin{equation}
Y^2=X^3-\frac{15}{16}X+\frac{11}{32} ~,
\qquad
\beta=\frac{1}{2}\,\left(1+\frac{X^2-X+\frac{7}{16}}{(X-\frac{1}{2})^2}\,Y\right)
~.
\end{equation}
Again, we can easily check, by finding the order of vanishing at the critical points of map, that the ramification structure is
$\left\{ \begin{array}{l} 3,3\\3,3\\3,3\end{array}\right\}$, which is indeed as expected for $\mathbb{C}^2/\mathbb{Z}_2\times \mathbb{C}$.

This concludes our treatment of the orbifolds of $\mathbb{C}^3$, and we see a consistent and enthralling story woven between unbranched covers of tori and the orbifolding construction of the Belyi pair.

\comment{
\subsection{Weierstra\ss\ identities}
The multi-angle formulas for the Weierstra\ss\ $\wp$-function enable us to consider other covers of the torus.
We have that:
\be
\wp(nz; \{g_2, g_3\}) = \wp(z, \{g_2, g_3\}) - \frac{\psi_{n+1} \psi_{n-1}}{\psi_n^2} ~,
\ee
where
\bea
\psi_1 &=& 1 ~, \nn \\
\psi_2 &=& -\wp'(z; \{g_2, g_3\}) ~, \nn \\
\psi_3 &=& 3\wp(z; \{g_2, g_3\})^4 - \frac32 g_2 \wp(z; \{g_2, g_3\})^2 - 3 g_3 \wp(z; \{g_2, g_3\}) - \frac{g_2^2}{16} ~, \nn \\
\psi_4 &=& \wp'(z; \{g_2, g_3\}) (-2\wp(z; \{g_2, g_3\})^6 + \frac{5g_2}{2} \wp(z; \{g_2, g_3\})^4 + 10 g_3 \wp(z; \{g_2, g_3\})^3 + \nn \\
&& \frac{5g_2^2}{8} \wp(z; \{g_2, g_3\})^2 + \frac{g_2 g_3}{2} \wp(z; \{g_2, g_3\}) + g_3^2 - \frac{g_2^3}{32} ) ~, \\
\psi_n &=& -\frac{1}{\wp'(z; \{g_2, g_3\})} \psi_{\frac{n}{2}} \left( \psi_{\frac{n}{2}+2} \psi_{\frac{n}{2}-1}^2 - \psi_{\frac{n}{2}-2} \psi_{\frac{n}{2}+1}^2 \right) ~, \frac{n}{2}\in \mathbb{Z} ~, \nn \\
\psi_n &=& \left( \psi_{\frac{n-1}{2}+2} \psi_{\frac{n-1}{2}}^3 - \psi_{\frac{n-1}{2}-1} \psi_{\frac{n-1}{2}+1}^3 \right) ~, \frac{n-1}{2}\in \mathbb{Z} ~. \nn
\eea

\todo{Perhaps include a discussion of multiangle formulas before moving on and $n$-fold covers before moving on.}
}

\section{The conifold and its orbifolds}
\label{Conifoldexample}\setall

To exhibit the general applicability of our methodology, we consider
other Calabi--Yaus.
The second most famous theory, after $\mathbb{C}^3$, in the AdS/CFT dictionary, is arguably the conifold theory.

From~\cite{Jejjala:2010vb} we recall that its Belyi pair is:
\begin{equation}
y^2=x^3-x ~, \qquad \beta=\frac{(x+1)^2}{4\,x} \ ,
\end{equation}
with ramification structure $\left\{ \begin{array}{l} 4\\4\\2,2\end{array}\right\}$.
In \textit{cit.~ibid.}, also the phase $I$ of the chiral $\mathbb{Z}_2$ orbifold of the conifold, namely $\mathbb{F}_0^I$, the first toric phase of the cone over the zeroth Hirzebruch surface, was found.
The corresponding pair was already presented in our case study in equation~(\ref{B-F0}), and we recall them here as:
\begin{equation}\label{F0-I}
y^2=x^3-x ~, \qquad
\beta_I=i\frac{(i+x)^4}{8\,x\,(1-x^2)} ~.
\end{equation}

It is easy to see that the following transformation takes the conifold pair into the $\mathbb{F}_0^I$ one:
\begin{equation}
\label{orbF0}
(x,\,y)\ \mapsto\ \left( \frac{2 i x}{-1 + x^2},\, \sqrt{2}\, e^{i3\pi/4} \, \frac{1+x^2}{(1-x^2)^2}\,y \right) ~.
\end{equation}
We can further check this transformation by noting that acting twice thereby produces a doubling of the periods, and thus we can apply and compare with the standard double angle formulae.
For the conifold case we have $g_2=1~, g_3=0$ so
\begin{equation}
x\ \mapsto\ \frac{1}{4}\, \frac{(x^2+1)^2}{x^3-x}
\end{equation}

On the other hand, if we act two times with our transformation, we find
\begin{equation}
x\ \mapsto\ 4\, \frac{x^3-x}{(x^2+1)^2}
\end{equation}
However, in the conifold case an automorphism of the pair is $x\mapsto 1/x$, so this transformation in fact coincides with the one coming from the double angle formulae.

\subsection{Covers and orbifolds at degree $2$ }

Let us now find the Belyi pairs for degree two covers of the orbifold by
using the method for explicit unbranched covers described
in Section~\ref{chicha}.

Our ansatz for the transformation is:
\begin{equation}
(x,\,y)\ =\ \left(\frac{X^2+\alpha_1\,X+\alpha_0}{X+\beta_0},\, \frac{X^2+2\,\beta_0\,X+(\alpha_1\,\beta_0-\alpha_0)}{(X+\beta_0)^2}\, Y\right) ~.
\end{equation}
Upon substitution into~(\ref{mastereq}) we find essentially two classes of non-trivial solutions below:

\begin{enumerate}
\item
\textbf{Recovering $\mathbb{F}_0$: }

The first such solution is
\begin{equation}
(x,\,y)\ =\ \left(\frac{4\,X^2+1}{4\,X},\,\frac{4\,X^2-1}{4\,X^2}\,Y\right)
\end{equation}
which takes the curve into $Y^2=X^3+\frac{X}{4}$.
Defining the coordinate change
\begin{equation}
(X,\,Y)\ \mapsto\ \left(\frac{i\,(1+X)}{2\,(X-1)},\,\frac{i-1}{2\,(X-1)^2}\,Y\right) ~,
\end{equation}
one can easily check that the curve goes to the desired $Y^2=X^3-X$ while the transformation becomes the expected result in~(\ref{orbF0}).

\item
\textbf{Another $\mathbb{Z}_2$ cover}

The solution yielding to $\mathbb{F}_0^I$ is not the only one. In fact, it is straightforward to check that the following transformation\footnote{There is yet another solution with the same properties as this one. It should be related to enlarging the unit cell along the other direction.} is also a solution of our general set of equations for degree two covers
\begin{equation}
(x,\,y)\ =\ \left(\frac{(1+4\,X)^2}{8\,(1+2\,X)},\,\frac{X^2+X+\frac{3}{16}}{(X+\frac{1}{2})^2} Y \right) ~.
\end{equation}
The resulting curve is $Y^2=X^3-\frac{11}{16}X-\frac{7}{32}$ and
it is immediate to check that $j=287496$.

On the other hand, this $\mathbb{Z}_2$ cover should correspond to enlarging the unit cell in one direction.
That should correspond to $\tau\rightarrow 2\,\tau$. The original curve had $j$-invariant 1728, which corresponds to $\tau=i$ and hence a square unit cell.
Thus, we should expect the enlarged unit cell $\mathbb{Z}_2$ cover to have $j=j(2\,i)$, which in fact is the expected 287496.

We have constructed the cover corresponding to a non-chiral $\mathbb{Z}_2$ orbifold of the conifold.\footnote{
Note that, as expected, when thought from a generic point of view as described in the first section, the construction of the cover makes no reference whatsoever to the Belyi function.}

We can now plug this into the original conifold Belyi map and find that for this $\mathbb{Z}_2$ orbifold and find the Belyi pair to be
\begin{equation}\label{L222-I}
Y^2=X^3-\frac{11}{16}-\frac{7}{32} ~, \qquad
\beta_{II}=\frac{(3+4\,X)^4}{32\,(1+2\,X)\,(1+4\,X)^2}~.
\end{equation}

After some computation, one can check that this is a Belyi map, with
ramification structure
$\left\{ \begin{array}{l} 4,4\\4,4\\2,2,2,2 \end{array}\right\}$.
This has the correct properties for $\mathcal{C}/\mathbb{Z}_2$, a space to which one is commonly referred as $L^{222}$.
In fact, the map above corresponds to the first toric phase of such a theory,
as we will discuss momentarily.
\end{enumerate}

\section{Seiberg duality and $ \tau_B , \tau_R $ }
\label{Seibergduality}\setall

As first noted in~\cite{Feng:2000mi,Feng:2001xr}, typically for a given Calabi--Yau singularity more than a single dual gauge theory can be found.
Such different ``phases,'' dubbed \textit{toric phases}, are in fact connected by Seiberg duality.
It is thus very natural to ask how Seiberg duality arises in the context of the Belyi construction.
To this end, we continue the $\mathbb{Z}_2$ examples of the conifold described in the last section and we show the Belyi pairs for the corresponding Seiberg dual phases.

\subsection{The two phases of $\mathbb{F}_0$}

It is by now well known that the cone over the zeroth Hirzebruch surface $\mathbb{F}_0$ has two toric Seiberg dual pairs, which we will denote by $\mathbb{F}_0 (I)$ and $\mathbb{F}_0 (II)$.
The Belyi pair for $\mathbb{F}_0(I)$ was already presented in~(\ref{F0-I}).
The second phase has ramification structure
$\left\{ \begin{array}{l} 3,3,3,3\\3,3,3,3\\2,2,4,4 \end{array}\right\}$, which conveniently falls into the category of the absence of unpaired odd indices, and can thus be readily treated by the $x$-only beta ansatz.
We find the Belyi pair to be
\begin{equation}\label{F0-II}
y^2=x^3-x ~, \qquad
\beta_{II}=i\,\frac{(x^2-(-1)^{1/3})^3}{3\,\sqrt{3}\,x^2\,(x^2-1)} ~.
\end{equation}

\begin{figure}[ht!]
\begin{center}
\includegraphics[scale=.7]{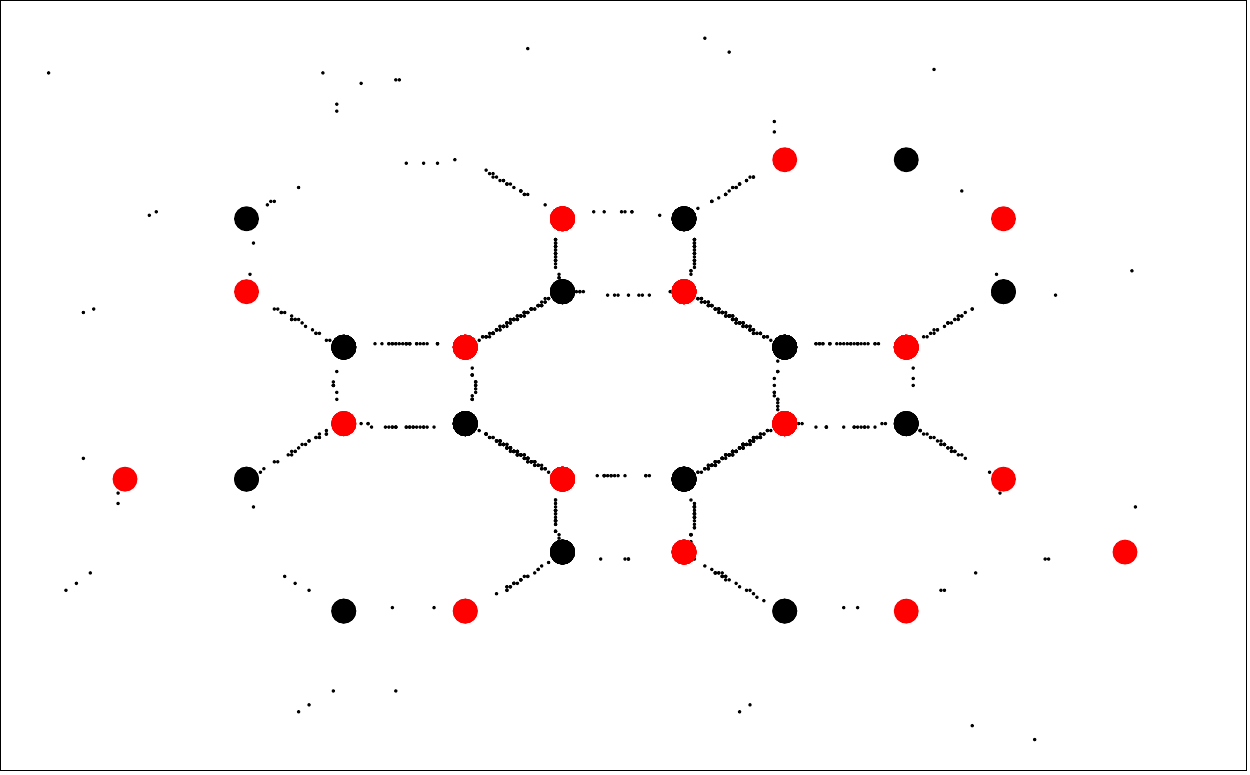}
\end{center}
\caption{{\sf \small Pre-image of the interval $[0,\,1]$, drawn by numerically inverting the Weierstra\ss\ function, explicitly recovers the $\mathbb{F}_0 (II)$ dimer from the Belyi pair in~(\ref{F0-II}).}}
\label{F0IInumeric}
\end{figure}

In fact, armed with the method of explicitly drawing the dimer from the Belyi map using the inverse Weierstra\ss\ function as expounded in~(\ref{curve}), we
 can readily check this map by drawing the pre-image of the $[0,\,1]$ segment, which in this case must be done fully numerically.
The result is shown in Figure~\ref{F0IInumeric}.

It is easy to construct and compute the $\tau_R$ of the isoradial dimer, which is $\tau_R=i$.
On the other hand, from the pair above it is easy to see that in fact $\tau_B=i$, thus also satisfying $\tau_B=\tau_R$.
We stress that this does not follow automatically, as phase II is not obtained by constructing a double cover of the parent theory, as is the case for phase I.

We also note that $\tau_R$ is the same in both Seiberg dual phases.
In fact, this is a general statement true for any pair of Seiberg dual theories, and will be addressed in a forthcoming work~\cite{Seibergdualtaus}.

\subsection{The two phases of $L^{222}$ and a counterexample to the $\tau_R=\tau_B$ conjecture}

The first phase of the non-chiral $\mathbb{Z}_2$ orbifold of the conifold $L^{222}$ was described in~(\ref{L222-I}).
There is also a Seiberg dual phase of the geometry.
For the dimer models of these theories, we again refer the reader to~\cite{Davey:2009bp}.
This second phase has the ramification data
$\left\{ \begin{array}{l} 3,3,4\\3,3,4\\2,2,3,3 \end{array}\right\}$ and once more luckily falls into the $x$-only beta ansatz.
We thus readily arrive at the quite non-trivial Belyi pair:
\begin{equation}
y^2 = x(x-1)(x - (8 (-20 - 9 \sqrt{5}))^{-1}) ~, \qquad
\beta = \frac{25 x^2 (-5 + 5 \sqrt{5} + 16 x)^3}{8 (-5 + 2 \sqrt{5} + 10 x)^3 (-20 + 9 \sqrt{5} + 40 x)} ~.
\label{eq:L222pair}
\end{equation}

\begin{figure}[h!t!]
\begin{center}
\includegraphics[scale=.7]{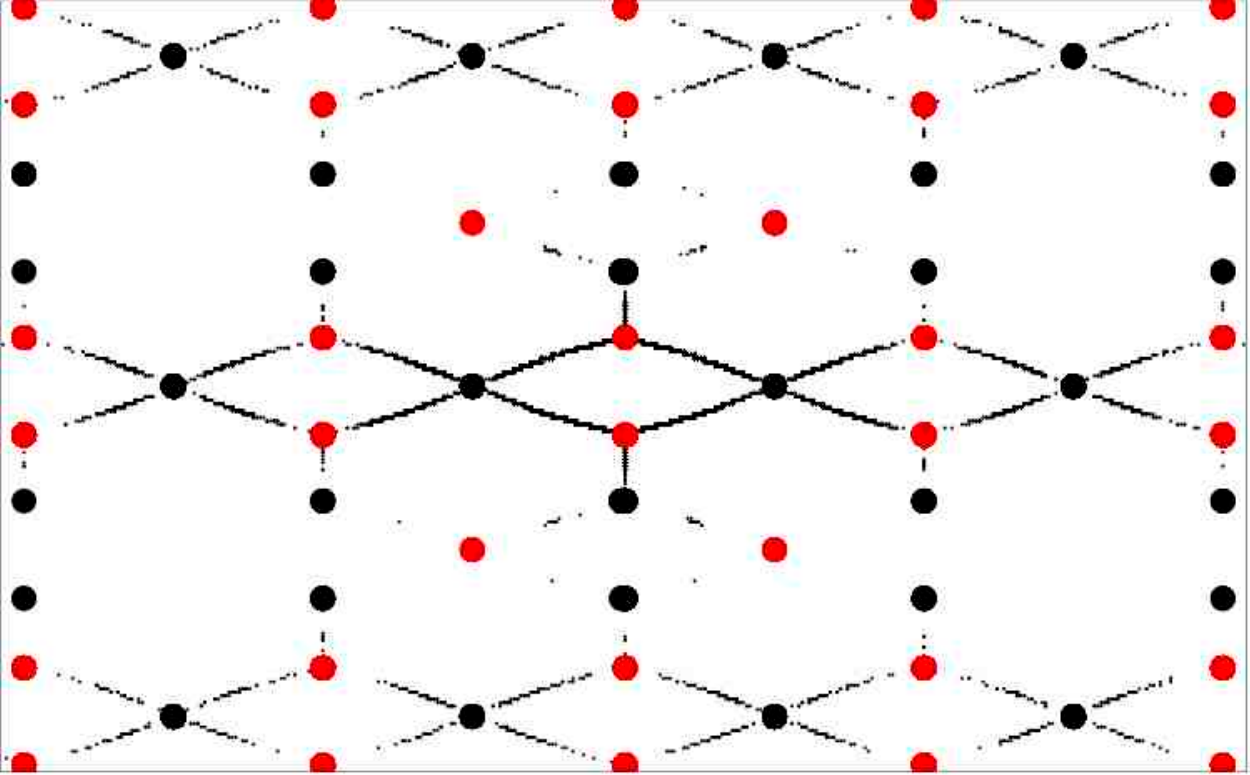}
\end{center}
\caption{{\sf \small
Pre-image of the interval $[0,\,1]$, drawn by numerically inverting the Weierstra\ss\ function, explicitly recovers the $L^{222} (II)$ dimer.}}
\label{L222IInumeric}
\end{figure}

Furthermore, we can numerically compute the pre-image of the segment $[0,\,1]$ recovering the $L^{222} (II)$ dimer, as shown in Figure~\ref{L222IInumeric}.
From the explicit expression of the curve we can compute the $j$-invariant, using the form in~(\ref{Clam}):
\begin{equation}
j= \left.\frac{256 (1 - \lambda + \lambda^2)^3}
{\lambda^2(1 - \lambda)^2} \right|_{\lambda = (8 (-20 - 9 \sqrt{5}))^{-1}}
= \frac{132304644}{5} ~.
\end{equation}

On the other hand, at the isoradial embedding, the dimer has $j=287496$.
As these two values of the $j$-invariant are different, this implies that in this case $\tau_R \ne \tau_B$, thus providing a counterexample to the conjecture in~\cite{Jejjala:2010vb}.
Nevertheless, we still note that $\tau_R$ is equal among the two Seiberg dual phases, which, as mentioned, holds universally~\cite{Seibergdualtaus}.

The database of~\cite{Davey:2009bp} contains three distinct
consistent dimers with the ramification structure of $L^{222}(II)$.
In Appendix~\ref{P1belyi} we recover this count from
enumerating permutation triples. We have only found one
such Belyi map (up to holomorphic reparameterizations of the $\mT^2$)
in the $x$-only ansatz, which we identified with $L^{222}(II)$
using the plot. After explaining
the connection of the $x$-only ansatz to
$\mP^1 \rightarrow \mP^1 $ Belyi maps in Appendix~\ref{P1belyi}, we count the
permutation triples for the latter, arriving at a unique equivalence
class, consistent with the analytic finding above.

\subsection{ $ \tau_B$ and $ \tau_R $ revisited }

In the case of $\mathbb{F}_0$ above, we have found one new example of $\tau_B = \tau_R$, which was not predicted by orbifolding, but which held in two phases related by Seiberg duality.
This led us to investigate the equality in the context of phases of $L^{222}$, where it did not hold.
Dimers, $R$-charges, and Belyi theory have led to an intriguing situation where the combinatoric data of the dimer leads to two complex structures on the torus.
For $\mC^3$, the conifold, and all their infinitely numerous orbifolds, the equality holds, but we have shown here that it is not completely general.
A general characterization of when $\tau_B = \tau_R$ and when it fails is an interesting problem.
An elliptic curve in a Belyi pair, when written in the parameterization~\eref{Clam}, has the property that $\lambda\in\bmQ$.
As the field $\bmQ$ is algebraically closed, this implies that $j(\lambda) = j(\tau_B)$ is itself algebraic.
Thus, if the conjecture in~\cite{Jejjala:2010vb} were true, then it would necessarily follow that $j(\tau_R)$ is algebraic as well.
Though we know that conjecture equating the complex structures does not generally hold, it may still nevertheless be true that $j(\tau_R)$ is algebraic.
The counterexample does not preclude this possibility.

The considerations here have motivated a proof that $\tau_R$ is invariant under toric (Seiberg) duality~\cite{Seibergdualtaus}.
Given the deep number theory behind Belyi pairs~\cite{grotesquisse}, this also raises many questions about the number theoretic structure of the $R$-charges in connection with toric duality, which we hope to address in the future.

\section{Conclusion}
\label{conclusions}\setall

Dimer models have proven to be a very efficient and deep way to encode SCFTs dual to D$3$-branes probing a toric Calabi--Yau conical singularity.
In turn, an elegant way of encoding the dimer into a Belyi pair was developed in~\cite{Jejjala:2010vb}.
The Belyi pair provides the torus with a complex structure $\tau_B$.
It was observed that the isoradial embedding of the dimer~\cite{Hanany:2005ss} provides another complex structure $\tau_R$ for the torus after $a$-maximization~\cite{Intriligator:2003jj}.
A general understanding of the relation between these two complex structures is not available, but infinitely many examples where they agree were found~\cite{Jejjala:2010vb}, which led to a conjecture that the agreement holds for all dimers.

The Belyi construction is very rigid, which in particular makes it very hard to find explicit examples of pairs.
In this paper, we have described computational methods for finding explicit examples in the case where the Belyi curve has genus one, which is relevant to dimer models for toric SCFTs.
We have described the combinatoric structure of a class of Belyi pairs, in terms of the ramifications of the Belyi map, which allows the reduction of the Belyi pair construction to a simpler one involving $\mP^1 \rightarrow \mP^1$ (Section~\ref{xonly} and Appendix~\ref{P1belyi}).
Substantial progress is also achieved (in Section~\ref{orbifolds} and Appendix~\ref{propweiss}) for the case of Belyi pairs for orbifold Calabi--Yaus.
Given the Belyi pair for a parent theory one can construct the pairs for its degree $n$ orbifold daughters by taking $n$-fold unbranched covers of the original torus~\cite{Jejjala:2010vb}.
In this note we have provided a systematic way for constructing such covers.
We have shown how the known examples are recovered as well as produced new pairs.

Degree $n$ covers are fully specified by a set of three numbers.
The methods described in this paper produce all degree $n$ covers, but a systematic identification of the precise cover for each specified set of three numbers, is not yet available.
Progress in this area will likely exploit developments on the counting of orbifolds of specified symmetry types~\cite{Hanany:2010cx, Davey:2010px, Hanany:2010ne, newpaper}.
We postpone these investigations for future work.

The explicit construction of the $n$-fold covers allows us to, in principle, construct the Belyi maps corresponding to $\frac{\mathbb{C}^2}{\mathbb{Z}_n}\times \mathbb{C}$, out of which we have explicitly shown the $n=2$ example.
As these theories are $\mathcal{N}=2$ supersymmetric, they are a natural arena possible connections to discussions of Belyi pairs in the context of Seiberg--Witten curves~\cite{Ashok:2006br}.

Beyond orbifolding, another natural field theory operation on dimers is Seiberg duality, which can generically produce new toric phases from orbifolds.
We have found the Belyi pairs for Seiberg dual phases of the $\mathbb{Z}_2$ orbifolds of the conifold --- both chiral, \textit{e.g.}, $\mathbb{F}_0$, and non-chiral, \textit{e.g.}, $L^{222}$ cases.
In the former case, Seiberg duality preserves $\tau_B, \tau_R$ hence the $\tau_B = \tau_R$ relation inherited from the orbifold.
In the case of $ L^{222} $ however, Seiberg duality preserves $\tau_R$ but not $\tau_B$, so we obtain a counterexample to the $\tau_R=\tau_B$ conjecture in~\cite{Jejjala:2010vb}
These models exemplify the general property that $\tau_R$ is preserved across Seiberg dual phases, a fact we prove in a forthcoming publication~\cite{Seibergdualtaus}.

\section*{Acknowledgements}
We are delighted to thank Rak-Kyeong Seong for discussions.
YHH is indebted to the gracious patronage of the Science and Technology Facilities Council, UK, for an Advanced Fellowship, the Chinese Ministry of Education, for a Chang-Jiang Chair Professorship at NanKai University, as well as City University, London and Merton College, Oxford, for their enduring support, \textit{et hoc die bibet ad regem reginamque futuri Angliae.}
YHH and VJ acknowledge NSF grant CCF-1048082.
VJ and SR are supported by an STFC grant ST/G000565/1.
DRG thanks the Theory Group at University of Oviedo for warm hospitality while this work was being finished.
DRG is supported by the Israel Science Foundation through grant 392/09.
He also acknowledges support from the Spanish Ministry of Science through the research grant FPA2009-07122 and Spanish Consolider-Ingenio 2010 Programme CPAN (CSD2007-00042).

\begin{appendix}

\section{Properties of the Weierstra\ss\ $\wp$-function}
\label{propweiss}

In this Appendix, we collect some of the properties of the Weierstra\ss\ $\wp$-function and explain how these can be used to generate torus covers and Belyi maps.
The Weierstra\ss\ $\wp$-function may be expressed a series:
\be
\wp(z;\omega_1,\omega_2)=\frac{1}{z^2}+ \sum_{(m,n) \ne (0,0)} \left\{ \frac{1}{(z-m\omega_1-n\omega_2)^2} - \frac{1}{\left(m\omega_1+n\omega_2\right)^2} \right\} ~.
\ee
The function has a double pole at $z=0$.
It is a meromorphic doubly periodic function of $z$, with periods $(2\omega_1, 2\omega_2)$.

Setting $x = \wp (z; \omega_1, \omega_2)$ and $y = \wp'(z; \omega_1, \omega_2)$, we have
the equation
\be
y^2 = 4 x^3 - g_2 x - g_3 = 4 (x- e_1) (x - e_2) (x - e_3) ~.
\ee
Thus, we have
\bea
0 &=& e_1 + e_2 + e_3 ~, \cr
g_2 & = & -4 (e_1 e_2 + e_1 e_3 + e_2 e_3) = 4 (e_1^2 + e_2^2 + e_1 e_2) ~, \cr
g_3 & = & 4 e_1 e_2 e_3 = - 4 e_1 e_2 (e_1 + e_2) ~.
\label{eq:g2g3}
\eea
The important point to note is that we can write the parameters $g_2$ and $g_3$ in the elliptic curve as functions of $e_1$ and $e_2$.

\paragraph{The double cover:}
The book by P.~du Val~\cite{duval} gives a beautiful presentation of some relevant identities which are ideally suited to constructing degree $d$ covers.
We will quote some formulae relating a double cover in terms of an elliptic curve written in the coordinates $(X, Y)$ with the parameters $(E_1, E_2, E_3; G_2, G_3; \omega_1, \omega_2)$ of an elliptic curve written in the coordinates $(x, y)$ with parameters
$(e_1, e_2, e_3; g_2, g_3; \frac12 \omega_1, \omega_2)$.
$X$ is a Weierstra\ss\ function obtained as a sum over the points of a lattice $\Omega$ generated by $(2 \omega_1, 2 \omega_2)$.
To get $x$, once simply sums $X(z) $ with $ X(z + \omega_1)$ and shifts by a constant:
\bea\label{sumX}
x(z) & = & X(z) + X(z + \omega_1) - E_1 \cr
&=& X + \frac{(E_1 - E_2)(E_1 - E_3)}{X - E_1} \cr
&=& \frac{X^2 - E_1 X + 3 E_1^2 - \frac14 G_2}{X - E_1} ~.
\eea
The second line follows by using an identify for shifts of
 Weierstra\ss\ functions by a half-period.
Then we differentiate to get
\bea
y(z) = \frac{X^2 - 2 E_1 X - 2 E_1^2 + \frac14 G_2}{(X - E_1)^2}\, Y ~.
\eea
Recall that $G_2$ is a function of $E_1$ and $E_2$, as in \eref{eq:g2g3}.
To relate the parameters of the elliptic curve corresponding to $\mathbb{T}^2$ to its cover, we make use of the following relations:
\be
g_2 = 60 E_1^2 - 4 G_2 ~, \qquad
g_3 = 14 G_2 E_1 + 22 G_3 ~.
\label{eq:solvethese1}
\ee
One checks explicitly that, with these relations, that the relation $y^2 = 4 x^3 + g_2 x - g_3$ reduces to $Y^2 = 4 X^3 - G_2 X - G_3$.

\paragraph{The triple cover:}
We shall again follow~\cite{duval} and define $\omega_{3,4} = \mp \omega_1 - \omega_2$.
The function $\wp(\frac23\omega_i)$ assumes the values of the roots of the quartic equation
\be
X^4 - \frac12 G_2(E_1, E_2) X^2 - G_3(E_1, E_2) X - \frac{1}{48} G_2(E_1, E_2)^2 = 0 ~.
\ee
Let us call these roots $A_1, \ldots, A_4$.
Note that the derivative of this expression equated to $Y^2$ gives the equation for the elliptic curve of the cover.

In terms of the coordinates of the cover, the original torus may be written using the variables
\bea
x &=& \frac{X^3 - 2 A_j X^2 + (7 A_j^2 - \frac12 G_2) X - (2A_j^3 + \frac12 G_2 A_j + G_3)}{(X-A_j)^2} ~, \\
y &=& \frac{X^3 - 3 A_j X^2 - (3 A_j^2 - \frac12 G_2) X - (3 A_j^3 - \frac32 G_2 A_j - 2 G_3)}{(X-A_j)^3}\, Y ~.
\eea
The parameters of the original torus and its cover are related as
\be
g_2 = 120 A_j^2 - 9 G_2 ~, \qquad
g_3 = 280 A_j^3 - 42 G_2 A_j - 27 G_3 ~.
\label{eq:solvethese2}
\ee

\paragraph{Constructing Belyi maps:}
The parameters of the elliptic curve $G_2$ and $G_3$ are known functions of $E_1$ and $E_2$ as stated in~\eref{eq:g2g3}.
Using this information, the $A_j$ are as well known functions of $E_1$ and $E_2$ as well.
Given $g_2$ and $g_3$ that determine the original torus, we can therefore solve \eref{eq:solvethese1} or \eref{eq:solvethese2}, at least numerically, for $E_1$ and $E_2$.
This enables us to write explicit functions $x(X,Y)$ and $y(X,Y)$.
The Belyi pair is then
\be
\left( Y^2 = 4 X^3 - G_2(E_1, E_2) X - G_3(E_1,E_2) ~,~ \widehat\beta(X,Y) = \beta(x(X,Y),y(X,Y)) \right) ~.
\ee

We have used this procedure to construct the Belyi pairs associated to $\mathbb{Z}_2$ and $\mathbb{Z}_3$ orbifolds of $\mathbb{C}^3$ and ${\cal C}$.
The procedure generalizes to covers of higher degree.

\section{The $x$-ansatz and $ \mP^1 \rightarrow \mP^1$
Belyi maps}\label{P1belyi}

The $x$-ansatz construction of Section~\ref{algorithm} starts with a Belyi curve
\bea\label{curveeq}
y^2 = x(x-1)(x-\lambda)
\eea
and a Belyi map
\bea\label{mapeq}
\beta(x,y) = {P(x) \over Q(x) } ~.
\eea
Let us define $b(x) = {P(x) \over Q(x)}$.
We can prove, using the discussion of local coordinates at and away from the points $\{0,1,\lambda,\infty\}$ that when the equations~(\ref{curveeq}) and~(\ref{mapeq}) give a Belyi pair, then $b(x)$ is a Belyi map from $\mP^1$ to $\mP^1$.
The proof proceeds, by showing that ramification points of $b$ away from $\{0,1,\lambda,\infty\}$ correspond to pairs (depending on choice of sign of $y$) of ramification points of $\beta$, and hence map to $\{0,1,\infty\}$ if $\beta$ is Belyi.
We say that the \textit{non-special} ramification points of $\beta$ arise from \textit{replicating} the ramification points of $b$.
The next step shows that the special points $\{0,1,\lambda,\infty\}$ are always ramification points of $\beta$, with twice the ramification index of $b$, hence must map to $\{0,1,\infty\}$.
We say that the \textit{special} ramification points of $b$ are \textit{doubled} in $\beta$.
This completes the proof.

Conversely, it is also true that any Belyi map $b~:~\mP^1\rightarrow \mP^1$ with more than three ramification points, provides, upon a choice of four of these points $\{x_1,x_2,x_3,x_4\}$, to a Belyi pair
\bea
y^2 = (x-x_1)(x-x_2)(x-x_3)(x-x_4) ~; \quad \beta(x,y) = b(x) ~.
\eea
A standard transformation of elliptic curves takes this to cubic form~\cite{wolfquartic}.
If we assume one of the four points is originally at infinity, then we immediately get a cubic Belyi curve.

We can apply this knowledge to show that the Belyi map describing phase two of $L^{222}$ is the unique map with the ramification structure
\be
\left\{ \begin{array}{l} 3,3,4\\3,3,4\\2,2,3,3 \end{array}\right\} ~.
\label{eq:newrams}
\ee
that falls into the $x$-only beta ansatz.
Let us first consider the cycle structure of permutations that could potentially yield a Belyi map with the data~\eref{eq:newrams}.
Suppose we fix
\be
\sigma_W = (1\,2\,3)(4\,5\,6)(7\,8\,9\,10) ~.
\ee
We can always do this by labeling the fields appropriately.
The lengths of the cycles correspond to the ramification indices.
We know from~\eref{relation} that
\be
\sigma_\infty = \sigma_W^{-1} \cdot \sigma_B^{-1} ~.
\ee
There are $598$ potential $\sigma_B\in S_{10}$ that have the cycle structure $\{3,3,4\}$ and yield a $\sigma_\infty$ with cycle structure $\{2,2,3,3\}$.
But the permutations are defined only up to conjugation equivalence.
Note that the automorphism group of ${\rm Aut}(\sigma_W)$ is generated as:
\be
\langle 1, (1\,2\,3), (4\,5\,6), (7\,8\,9\,10), (1\,4)(2\,5)(3\,6) \rangle ~.
\ee
There are $72$ elements $\gamma\in {\rm Aut}(\sigma_W)$, and the equivalence $\sigma_B' \sim \gamma \sigma_B \gamma^{-1}$ reduces the $598$ to twelve permutations.
Using the prescription of~\cite{Jejjala:2010vb} to write the zig-zag paths from the permutations and dropping inconsistent dimers, for which the zig-zag paths self-intersect, as well as disconnected dimers, there are three potential $\sigma_B$:
\be
\sigma_B \in \left\{ (1\,2\,4)(3\,7\,8)(5\,6\,9\,10), (1\,2\,4\,5)(3\,7\,8)(6\,9\,10), (1\,4\,10)(2\,5\,8)(3\,7\,6\,9) \right\} ~.
\ee
We identify the connectivity of the corresponding bipartite graphs with the dimers of $L^{131}$, $L^{222}$ (II), and $dP_1$, respectively.
These are the only dimers in~\cite{Davey:2009bp} with the ramification data on which this analysis is based.

For one of these dimers to have a Belyi pair under the $x$-only ansatz, it must arise from a bipartite graph on $\mP^1$.
Examining~\eref{eq:newrams}, the dessin on $\mP^1$ should have the ramification data
\be
\left\{ \begin{array}{l} 3,2\\3,2\\1,1,3 \end{array}\right\} ~,
\label{eq:oldrams}
\ee
as the even ramification indices in~\eref{eq:newrams} arise from the doubling of a ramification index in~\eref{eq:oldrams} and the paired odd ramification indices in $\beta$ come from replicating.
Now, we can repeat the permutation analysis using elements of $S_5$.
Fixing
\be
\sigma_W = (1\,2)(3\,4\,5) ~,
\ee
and searching for $\sigma_B$ and $\sigma_\infty$ with the ramification structure quoted in~\eref{eq:oldrams}, we find that
\be
\sigma_B \in \left\{ (1\,2)(3\,4\,5), (1\,2\,5)(3\,4) \right\} ~.
\ee
The former possibility leads to a disconnected graph, so the latter is the appropriate choice.
This establishes that there is a single $\beta$ in the $x$-only ansatz that reproduces the ramification data in~\eref{eq:newrams}.
This is the map we have constructed in~\eref{eq:L222pair}.

\end{appendix}

\bibliography{beta} 
\bibliographystyle{utphys} 

\end{document}